\documentclass{emulateapj}

\usepackage{lscape}   % for emulateapj instead of rotate
\usepackage{subfig}

\def\Chandra{{\em Chandra}}

\newcommand{\aox}{\ifmmode{\alpha_{\mathrm{ox}}} \else $\alpha_{\mathrm{ox}}$\fi} 
\newcommand{\atoms}{\ifmmode{\mathrm{\,atoms~cm^{-2}}} \else \,atoms cm$^{-2}$\fi}
\newcommand{\ax}{\ifmmode{\alpha_x} \else $\alpha_x$\fi} 
\newcommand{\cmsq}{\ifmmode{\mathrm{cm^{-2}}} \else cm$^{-2}$\fi}
\newcommand{\degs}{\ifmmode ^{\circ}\else$^{\circ}$\fi}
\newcommand{\degsq}{\ifmmode {\mathrm{deg^2}} \else deg$^2$\fi}
\newcommand{\perdegsq}{\ifmmode {\mathrm{deg^{-2}}} \else deg$^{-2}$\fi}
\newcommand{\ew}{\ifmmode{W_{\lambda}} \else $W_{\lambda}$\fi}
\newcommand{\fcgs}{\ifmmode erg~cm^{-2}~s^{-1[B}\else erg~cm$^{-2}$~s$^{-1}$\fi}
\newcommand{\fnucgs}{\ifmmode {\mathrm{erg~cm^{-2}~s^{-1}~Hz^{-1}}}\else erg~cm$^{-2}$~s$^{-1}$~Hz$^{-1}$\fi}
\newcommand{\gax }{{\lower0.8ex\hbox{$\buildrel >\over\sim$}}}

\newcommand\ha{\ifmmode {\mathrm H}\alpha \else H$\alpha$\fi}
\newcommand\hb{\ifmmode {\mathrm H}\beta \else H$\beta$\fi}
\newcommand{\kms}{\ifmmode~{\mathrm{km~s}}^{-1}\else ~km~s$^{-1}~$\fi}
\newcommand{\lax }{{\lower0.8ex\hbox{$\buildrel <\over\sim$}}}
\newcommand{\lcgs}{\ifmmode erg~s^{-1}\else erg~s$^{-1}$\fi}
\newcommand{\lnucgs}{\ifmmode erg~s^{-1}~Hz^{-1}\else erg~s$^{-1}$~Hz$^{-1}$\fi}

\newcommand{\logz}{\ifmmode{\mathrm{log}}~z \else log$~z$\fi}
\newcommand{\lo}{\ifmmode l_o \else $~l_o$\fi}
\newcommand{\lx}{\ifmmode l_x \else $~l_x$\fi}
\newcommand{\lbol}{\ifmmode L_{\mathrm bol} \else $L_{\mathrm bol}$\fi}
\newcommand{\LEdd}{\ifmmode L_{\mathrm Edd} \else $L_{\mathrm Edd}$\fi}
\newcommand{\rEdd}{\ifmmode L/L_{\mathrm Edd} \else $L/L_{\mathrm Edd}$\fi}
\newcommand{\mbh}{\ifmmode {M_{\rm BH}}\else${M_{\rm BH}}$\fi}
\newcommand{\mdot}{\ifmmode \dot{m} \else $\dot{m}$\fi}
\newcommand{\mdote}{\ifmmode \dot{m}_{E} \else $\dot{m}_{E}$\fi}
\newcommand{\mone}{\ifmmode ^{-1}\else$^{-1}$\fi}
\newcommand{\msun}{\ifmmode {M_{\odot}}\else${M_{\odot}}$\fi}
\newcommand{\mtwo}{\ifmmode ^{-2}\else$^{-2}$\fi}
\newcommand{\nhgal}{\ifmmode{ N_{H}^{Gal}} \else N$_{H}^{Gal}$\fi}
\newcommand{\nh}{\ifmmode{\mathrm N_{H}} \else N$_{H}$\fi}
\newcommand{\nhintr}{\ifmmode{ N_{H}^{intr}} \else N$_{H}^{intr}$\fi}
\newcommand{\nhtot}{\ifmmode{ N_{H}^{tot}} \else N$_{H}^{tot}$\fi}
\newcommand{\nhz}{\ifmmode{ N_{H}^z} \else N$_{H}^z$\fi}
\newcommand{\oi}{\ifmmode{\mathrm [O\,II]} \else [O\,II]\fi}
\newcommand{\oii}{\ifmmode{\mathrm [O\,II]} \else [O\,II]\fi}
\newcommand{\oiii}{\ifmmode{\mathrm [O\,III]} \else [O\,III]\fi}
\newcommand{\optebl}{\ifmmode L_{\mathrm 2500\,\AA} \else $~L_{\mathrm 2500\,\AA}$\fi}
\newcommand{\opteml}{\ifmmode l_{\mathrm{2500\,\AA}} \else $~l_{\mathrm{2500\,\AA}}$\fi}
\newcommand{\xebl}{\ifmmode L_X \else $~L_X$\fi}
\newcommand{\xeml}{\ifmmode l_{\mathrm{2\,keV}} \else $~l_{\mathrm{2\,keV}}$\fi}

\def\geqsim{\lower.73ex\hbox{$\sim$}\llap{\raise.4ex\hbox{$>$}}$\,$}
\def\leqsim{\lower.73ex\hbox{$\sim$}\llap{\raise.4ex\hbox{$<$}}$\,$}

\shorttitle{Binary Quasars and Their Environments} 
\shortauthors{Green et al.}

\received{2011 May}
\begin{document}


\title{A Multiwavelength Study of Binary Quasars and Their Environments}


\author{Paul J. Green\altaffilmark{1,2}}
\email{pgreen@cfa.harvard.edu}
\author{Adam~D.~Myers\altaffilmark{2,3},
Wayne~A.~Barkhouse\altaffilmark{2,4},
Thomas~L. Aldcroft\altaffilmark{1},
Markos Trichas \altaffilmark{1}
Gordon T. Richards\altaffilmark{5}
\& \'Angel Ruiz \altaffilmark{6} \&
Philip F. Hopkins \altaffilmark{7}
}
\altaffiltext{1}{Harvard-Smithsonian Center for Astrophysics, 60 Garden
 Street, Cambridge MA 02138}
\altaffiltext{2}{Visiting Astronomer, Kitt Peak National Observatory
  and Cerro Tololo Inter-American Observatory, National Optical
  Astronomy Observatory, which is operated by the Association of
  Universities for Research in Astronomy, Inc. (AURA) under cooperative
  agreement with the National Science Foundation.}
\altaffiltext{3}{Department of Physics and Astronomy, University of Wyoming, 
  Laramie, WY 82071, USA}
\altaffiltext{4}{Department of Physics and Astrophysics, University of
  North Dakota, Grand Forks, ND 58202}
\altaffiltext{5}{Department of Physics, Drexel University, 3141
  Chestnut Street, Philadelphia, PA 15260, USA}
\altaffiltext{6}{Instituto de Física de Cantabria (IFCA), CSIC-UC,
  Avda. de los Castros, 39005 Santander, Spain}
\altaffiltext{7}{Department of Astronomy and Theoretical Astrophysics Centre,
University of California Berkeley, Berkeley, CA 94720, USA}

\begin{abstract}
We present\Chandra\ X-ray imaging and spectroscopy for 14 quasars in
spatially resolved pairs targeted as part of a complete sample of
binary quasars with small transverse separations drawn from Sloan
Digital Sky Survey (DR6) photometry.  We measure the X-ray properties
of all 14 QSOs, and study the distribution of X-ray and
optical-to-X-ray power-law indices in these binary quasars.  We find
no significant difference when compared with large control samples of
isolated quasars, true even for SDSS\,J1254+0846, discussed in detail
in a companion paper, which clearly inhabits an ongoing,
pre-coalescence galaxy merger showing obvious tidal tails. We present
infrared photometry from our observations with SWIRC at the MMT, and
from the WISE Preliminary Data Release, and fit simple spectral energy
distributions to all 14 QSOs.  We find preliminary evidence that
substantial contributions from star formation are required, but
possibly no more so than for isolated X-ray-detected QSOs.  Sensitive
searches of the X-ray images for extended emission, and the optical
images for optical galaxy excess show that these binary QSOs ---
expected to occur in strong peaks of the dark matter distribution ---
are not preferentially found in rich cluster environments.  While
larger binary QSO samples with richer far-IR and sub-millimeter
multiwavelength data might better reveal signatures of merging and
triggering, optical color-selection of QSO pairs may be biased against
such signatures.  X-ray and/or variability selection of QSO pairs,
while challenging, should be attempted.  We present in our Appendix a
primer on X-ray flux and luminosity calculations.
\end{abstract}



%\tableofcontents

% INTRODUCTION
\section{Introduction}
\label{sec:intro}

Supermassive black holes (SMBH; $\geqsim\,10^6\,\msun$) mainly grow from gas
accretion \citep{Soltan82,Merloni08} in the cores of active galaxies.
Luminous quasars, which host SMBH up to about a billion solar masses, are
already in place by redshift z$\sim$7 when the universe was much less
than a billion years old \citep{Fan01,Fan06,Mortlock11}.  While for
decades SMBH accretion was thought of largely as a product of galaxy
evolution, it is also seen now as a principal driver of that
evolution, based in large 
part on the surprisingly tight correlation between the mass of SMBH
and the mass or velocity dispersion of their host galaxy bulges
\citep{Ferrarese00,Gultekin09}.  If SMBH are the product of a sequence
of galaxy merger episodes (e.g., \citealt{Hernquist89, Kauffmann00,
 Hopkins08}), then binary SMBH are an inevitable outcome of galaxy
assembly.  As we seek greater understanding of the cosmic co-evolution
of galaxies and SMBH, the relatively rare binary SMBH hold unique
interest and promise, and particularly so for the most luminous
examples. 

% RARE PEAKS
Luminous quasars have always inhabited a relatively small fraction of
galaxies.  Studies of the clustering properties of quasars (two-point
correlation functions) indicate that the bias of quasars relative to
underlying dark matter increases rapidly with redshift, implying that
quasars inhabit rare massive dark matter halos of similar mass
($M_{\rm halo}\,\geqsim\,10^{12}\,h\mone \msun$) at every cosmic epoch
\citep{Porciani04,Croom05,Myers06,Myers07a,Shen07,Ross09}.  Though quasars
with a close ($<1\,$Mpc) quasar companion at comparable optical
luminosity constitute only $\sim0.1\%$ of optically selected quasars
overall, that represents a strong excess at small separations over the
extrapolation from the larger scale QSO spatial correlation function
\citep{Hennawi06,Myers08,Hennawi10,Shen10a}.  Indeed, the surprisingly
large number of binary quasars in the universe \citep{Djorgovski87,
  Myers07b,Hennawi06} is a key underpinning of the merger hypothesis.
Multiple authors \citep{Djorgovski91,Kochanek99, Mortlock99, Myers07b}
have noted that an excess of binary quasars could be due to tidal
forces in dissipative mergers that trigger inflow of gas towards the
nuclear region---and hence strong accretion activity in the nuclei of
merging galaxies.  However, this picture remains incomplete; are
mergers the {\em cause} of the observed excess of binary quasars or
rather is the excess of binary quasars the {\em result} of enhanced
small-scale clustering for the merger-prone halos that host quasars?
The measured small-scale excess including binary quasars ($R\ \geqsim
100$\,kpc) may {\em not} be due to mutual triggering, but rather
simply a statistically predictable consequence of overdense,
group-scale environments \citep{Hopkins08}.  These controversies
motivate detailed studies of binary quasars---and binary Active
Galactic Nuclei (AGN) in general.

% KPC-SCALE
At high redshifts, where the merging process is likely to be efficient
(e.g., \citealt{Springel05}), binary AGN are difficult to resolve.  At
more recent epochs, where they could be resolved, the merger rate is
lower \citep{Hopkins08}.  Nearby examples exist, however.  The
merger hypothesis is supported by (1) the existence of
spatially-resolved binary active galactic nuclei in a few $z<0.1$
galaxies with one or both of the nuclei heavily obscured in X-rays
(NGC\,6240; \citealt{Komossa03}; Arp\,299, \citealt{Zezas03};
Mrk\,463, \citealt{Bianchi08}), by (2) the unusual BL~Lac-type object
OJ~287 \citep{Sillanpaa88,Valtonen11} whose binary nature
is still under considerable debate \citep{Villforth10}, and perhaps by
(3) X-shaped morphology in radio galaxies (e.g., \citealt{Merritt02,
Liu04,Cheung07}). In addition, CXOC\,J100043.1+020637 contains two
AGN resolved at 0.5\arcsec\, ($\sim$2.5\,kpc) separation in HST/ACS
imaging, which have a radial velocity difference of
$\Delta\,v=150$~km/s, and appear to be hosted by a galaxy with a tidal
tail \citep{Comerford09,Civano10}.  

% SUB-KPC BINARY AGN
A recent flurry of searches for candidate close binary
AGN (with sub-kpc projected separations) has mostly involved
spectroscopic (unresolved) binaries.  Some show both broad and narrow
emission lines with significant velocity offsets, such as
SDSS\,J153636.22+1044127.0 \citep{Boroson09} or SDSS J105041.35+345631.3
\citep{Shields09}. Some may be true binary SMBH.  
SDSS\,J092712.65+294344.0 may be a binary SMBH with a single
disk \citep{Bog09,Dotti09}, or a single SMBH that has been ``kicked''
due to anisotropic emission of gravitational radiation near
coalescence \citep{Komossa08a}. 
Some may be similar to spatially unresolved quasars with
double-peaked broad emission lines (e.g., \citealt{Strateva03}).
Debate surrounding the various interpretations persists (e.g.,
\citealt{Lauer09, Wrobel09, Tang09, Vivek09, Chornock10}).  Many
spectroscopic binary AGN candidates with narrow emission lines only
have been selected from the Sloan Digital Sky Survey (SDSS;
\citealt{York00}) based on double-peaked \oiii\,
$\lambda\lambda$4959, 5007 emission lines in their fiber spectra
\citep{Wang09,Smith10,Liu10}.  Some remarkable objects have been found
(e.g., \citealt{Xu09}), but several scenarios can produce
double-peaked narrow emission lines, including projection
effects, outflows, jet-cloud interactions, special narrow-line region
(NLR) geometries, or even a merger where one AGN illuminates two
NLRs.  Near-infrared (near-IR) imaging and optical slit spectroscopy
can reveal genuine double-nuclei \citep{Liu10}, which constitute only
about 10\% of the candidates \citep{Shen10b}.  \Chandra\ imaging is
underway now to confirm these nuclei as AGN by resolving their luminous
hard X-ray emission.  \citet{Fu11} imaged 50 double-peaked
[O\,III]$\lambda$5007 AGN from the SDSS with Keck-II laser guide star
adaptive optics, confirming that most (70\%) are probably single AGN. 

Spatially resolved, confirmed mergers of broad line AGN may well be
the most useful systems for tracing the physics of the early merger
process because they probe {\em ongoing} mergers, and because the
spatial and velocity information, especially when combined with
well-resolved spectra providing separate black hole mass estimates,
offer more constraints on the properties of the merging components and
the evolution of the merger.  Examples of resolved binary broad line
AGN in confirmed mergers are virtually unknown.  Probably the best
example to date is SDSS\,J1254+0846 \citep{Green10}, which clearly
shows tidal tails from the ongoing merger.  Spatially-resolved active
binary mergers such as these provide by far the strongest constraints
on merger physics at kiloparsec scales.  Even when such obvious merger
signatures are not available, other probes of the properties of binary
quasars such as their environments, spectral energy distributions
(SEDs), and nuclear and host galaxy properties provide useful
information to help distinguish which systems may be undergoing merging or
triggering, and to elucidate merger physics itself.

% FURTHER PROBES
In this paper, we probe the multiwavelength properties of a small but
uniform sample of binary quasars, described in \S\,\ref{sec:sample}.
Using \Chandra\ X-ray imaging, in \S\,\ref{sec:cxo} we study the
high-energy SEDs of binary quasars (\S\,\ref{sec:cxoqso}) and how they
compare (in \S\,\ref{sec:compare}) to a subset of QSOs imaged in X-rays
by \Chandra\, as part of the \Chandra\ Multiwavelength Project
\citep{Green04,Green09}.
In \S\,\ref{sec:nirim} we present deep IR imaging we 
obtained at Mt Hopkins using the SAO Wide-field InfraRed Camera
(SWIRC) on the 6.5m MMT, to further examine the SEDs of binary QSOs.
Template-fitting to our multiwavelength SEDs presented in
\S\,\ref{sec:SEDs} will allow us to contrast binary QSOs directly with a
large sample of isolated QSOs from the
\Chandra\ Multiwavelength Project (ChaMP) in an upcoming paper.  

Binary quasars are expected to frequent massive dark matter halos,
which we test in \S\,\ref{sec:clusters}. We look for evidence of any
local hot intra-cluster medium (ICM) indicating a host group or
cluster in \S\,\ref{sec:cxocluster}.  X-ray cluster detection avoids
some of the pitfalls of optical/IR selection---namely, projection effects and
red-sequence bias towards evolved galaxy populations.  Our X-ray
imaging is sensitive even to poor clusters and groups with high $M/L$
\citep{Barkhouse06}, despite the presence of bright quasar point
sources \citep{Green05}.
Analysis of optical images of these fields is described in
\S\,\ref{sec:optim}.  We obtained NOAO/4m-MOSAIC images on Kitt Peak to 
deeply image the quasars (\S\,\ref{sec:optim4m}) in a search
for signs of merger activity or optical galaxy overdensities
associated with a host cluster, but SDSS imaging (\S\,\ref{sec:optimsdss})
largely turns out to be more useful than what we could obtain in variable
weather. 
We conclude with a brief discussion of our findings in \S\,\ref{sec:discuss}. 

Throughout, we assume a cosmology with $\Omega_{\rm m}=0.3,
\Omega_{\Lambda}=0.7$, and $H_0$ = 72  \kms\,Mpc$^{-1}$.

\section{Binary Quasar Sample}
\label{sec:sample}

Ongoing mergers hosting two luminous AGN are rare, so while a handful
of serendipitous examples exist, huge volumes of sky must be searched
to find them systematically. The SDSS provides a large sample of
multicolor imaging and  spectroscopy for this purpose.

Candidates in our sample (see Table~\ref{tsample}) were drawn from the
photometrically classified Type~1 (broad line) quasar catalog of
\citet{Richards09}.  All pairs of objects with component separations
of $2.9\arcsec$ to $6\arcsec$, ``UVX" or ``low-z" classification
flags \footnote{$\textbf{uvxts=1}$ OR $\textbf{lowzts=1}$} set and $g
< 20.85$ were targeted for follow-up spectroscopy. Selecting ``low-z"
and ``UVX" objects produces a sample that populates the redshift range
$0.4 < z < 2.4$.  At angular separations of $<2.9\arcsec$ we
supplemented our sample with SDSS\, J0740+2926 from \citet{Hennawi06}.

Our relatively uniform parent sample allows us to place these systems
in their larger cosmological context, which is crucial if we are to
understand the role of merger-triggered supermassive black hole
accretion, and its relationship to galaxy evolution.  By selection,
the two components of these quasar pairs are likely to have a wide
projection on the sky, which makes them useful for providing
morphological constraints on merger models.

As outlined in Table~\ref{tfits}, the quasar pairs we consider in this
work are confirmed, through spectroscopy, to be broad line quasar
pairs with components that are proximate in velocity space (``binary
quasars").  Five of our pairs had one object already confirmed
spectroscopically by SDSS.  (The $\sim$55\arcsec\,
minimum SDSS fiber separation usually precludes SDSS spectra for
both members.) The majority of the confirmed binary
quasars we study were followed up and spectroscopically identified in
previous works \citep{Hennawi06,Myers08}.  Exceptions are SDSS\,
J0813+5416 and SDSS\, J1254+0846.  These were spectroscopically
confirmed on February 9--11, 2008 at Kitt Peak National Observatory
using the R-C spectrograph on the Mayall 4-m telescope and the KPC-10A
grating. A $1.5\arcsec$ wide long-slit was oriented at a position
angle to simultaneously observe both candidates, and 20 minutes of
exposure time was sufficient to identify the candidates as broad line
quasars (the faintest quasar in these two binaries is at
$g=20.3$). Spectra were reduced using the
LowRedux\footnote{http://www.ucolick.org/$\sim$xavier/LowRedux/}
package.  better quality spectra of the QSOs in SDSS\, J1254+0846 were
obtained on 22 May, 2009 with Magellan/IMACS as detailed in
\citet{Green10}.

For Chandra observations, we restricted our sample to binary quasars
with velocity differences $\Delta v < 800$\,km\,s$^{-1}$, proper
transverse separations $R_{p}<30$\,kpc and redshifts $z < 1$.  As we
do not concern ourselves with binary quasars that do not meet these
criteria in this paper, any such candidates can be considered as
``discarded" for the purposes of this work, although they will be
published in a later paper (Myers et al., in preparation). Our
separation criterion selects hosts likely to be interacting on their
first or second pass. The velocity criterion removes most chance
projections but still allows for hosts in a variety of environments
from isolated pairs to massive clusters. The redshift criterion
prevents the necessary exposure times from becoming excessive. The
properties of our final sample are shown in Tables \ref{tsample} and
~\ref{tfits}.

\section{\Chandra\ X-ray Observations}
\label{sec:cxo}

% CHANDRA IMAGING
We obtained X-ray images of the seven quasar pairs with the \Chandra\ X-ray
Observatory on the dates shown in Table~\ref{tsample}.  We placed
targets near the ACIS-S aimpoint, and tuned our exposure times
to achieve $\sim$100~counts for the fainter member of each pair,
by converting the SDSS $r$ mag to an expected $f_X$ using the 75th
percentile X-ray-faintest value of log\,($f_X/f_r$)=$-0.5$ from the
\Chandra\ Multiwavelength Project (ChaMP) QSO sample \citep{Green09}.
For every pair, we convert $f_X$ to ACIS-S counts/sec using PIMMS,
with $\Gamma=1.9$ through \nhgal, and derive the exposure, which
yielded exposure times from 12 to 30\,ksec, with a total of
$\sim$172\,ksec.

\subsection{X-rays from the Quasars}
\label{sec:cxoqso}

The small (2--3\arcsec) separation of 3 of these pairs is not a
challenge for \Chandra.  In all cases but one, the X-ray 
components are detected, well-resolved by \Chandra, and correspond
closely ($<0.2\arcsec$) to their SDSS counterparts.  
SDSS\,J160602.81+290048.7 was not detected using {\tt wavdetect} 
and a detection significance threshold corresponding to
about one false source per ACIS chip.  However, aperture photometry 
at the optical source position shows 6 net counts, all above 2\,keV. 
To avoid cross-contamination between QSOs in each pair, we extracted
the X-ray photons from apertures 
corresponding to 90\% of the counts (for 1.5\,keV).  Some of the QSOs
in our sample yielded relatively few net counts.  In such cases,
instrumental hardness (photon count) ratios are often used, in the
belief that genuine spectral fitting is not warranted by the data
quality.  There are several problems with the use of hardness ratios (HRs).
HRs do not take redshift or intervening Galactic hydrogen column into
account.  They are difficult to interpret or compare, since they
convolve the intrinsic quasar SED with the telescope and instrument 
response, especially since the  latter depends on both time and ACIS
chip position\footnote{See information on ACIS contamination
at http://cxc.harvard.edu/cal}.  Finally, HRs waste useful
spectral information by crudely binning counts.  A direct spectral fit
of the counts distribution using the full instrument calibration,
known redshift, and Galactic column $N_H$ provides a much 
more direct measurement of quasar properties, most useful for
comparison to other quasar samples.  Correct spectral fitting does not
underestimate errors; even in the low-count regime, one can obtain
robust estimates of fit parameter uncertainties using the Cash (1979)
fit statistic.  The spectral fitting we employ provides the
most consistent and robust estimates of the physical parameters of 
interest---the power law slope and intrinsic absorption.

We fit an X-ray power-law spectral model  
 $$ N(E) = N(E_0)\, 
\Bigl(\frac{E}{E_0}\Bigr)^{(1-\Gamma)}\, \exp [-N_H^{Gal}\sigma(E)
              - N_H^z\sigma(E(1+z)) ] $$ 
%ADM maybe \sigma(E_{1+z}) is better. I'm not sure this is a double-function call? 
%ADM if changed it needs changed in the appendix, too.

\noindent to the counts for each QSO using the CIAO tool {\it Sherpa}, where
$N(E_0)$ is the  
normalization in photons cm$^{-−2}$ sec$^{-−1}$ keV$^{−-1}$ at
a reference energy $E_0$ (of 1~keV here), and $\sigma(E)$ is the absorption
cross-section \citep{Morrison83,Wilms00}.  We fix $N_H^{Gal}$  at the  
appropriate Galactic neutral absorbing column taken from
\citet{Dickey90}, and perform (1) a simple power-law fit
with no intrinsic absorption component (model $PL$)
and (2) include an intrinsic absorber with neutral
column $N_H^z$ at the source redshift  (model $PL_{\rm Abs}$). Unbinned
spectra were fit using Cash statistics  \citep{Cash79}. The best-fit
model parameters for all components are shown in Table~\ref{tfits}.  

The power-law energy index values  $\Gamma$ we measure are
typical of SDSS Type~1 (broad line) QSOs in general \citep{Green09},
with a mean of $2.14 \pm 0.29$ and a median of 2.11.
Unabsorbed fluxes and luminosities are calculated as detailed
in the Appendix, using the $\Gamma$ values from the $PL_{\rm Abs}$ fits
in every case except for the faintest object
SDSS\,J160602.81+290048.7, where we assume $\Gamma=2.1$. These values
only differ substantially from the $PL$-only fit values in the two
cases where there is absorption detected at $>$68\% confidence.   

\subsection{Comparison to Single Quasars}
\label{sec:compare}

Our small binary QSO sample has mean (median) redshift $0.72\pm0.18$
(0.77).  An excellent control sample is available already through the
ChaMP: we have matched 1175 SDSS QSOs from the SDSS photometric quasar
catalog \citep{Richards09} to \Chandra\ serendipitous X-ray sources
measured in 323 X-ray images from Cycles 1--6. To form a fair,
high-quality comparison sample, we limit the ChaMP QSOs to those at
$z<1.2$, with exposure times $>4$\,ksec, and off-axis angles
$\theta<12\arcmin$.  This yields a control sample of 264 isolated
QSOs, with mean (median) redshift $0.74\pm0.32$ (0.79), and a
cumulative \Chandra\ exposure of $\sim7.7$\,Msec.  The ChaMP sample
includes 70 QSO candidates (26\%) with only photometric redshifts
\citep{Weinstein04}. Since quasars are known to have significantly
higher $f_X/f_{\rm opt}$ values compared to galaxies (e.g.,
\citealt{Green04}), the additional criterion of X-ray detection 
for these photometric QSO candidates means that $\sim$98\% of them
are indeed QSOs, as found in \citet{Green09}.  Note that the parent
SDSS QSOs were selected optically using the same approach as we used  
to target binary quasars.  However, since we
specifically targeted the binary QSOs with \Chandra, their X-ray data
is of somewhat higher quality (all on-axis, with a mean/median of 573/319
X-ray counts) compared to the control sample (135/42 counts in the mean).
We find no significant difference in any of the measured ensemble
properties.  Comparing power-law fits, the mean (median) $\Gamma$\, is
$2.14\pm0.30$ (2.11) for the binary QSOs, and $1.96\pm0.61$ (2.01) for
the comparison sample. 

To compare X-ray/optical luminosity ratios, 
as in \citet{Green09}, we first estimate the monochromatic
luminosity at 2500\AA by finding the filter for each QSO whose
de-redshifted effective wavelength centroid (taken from
\citet{Fukugita96}) is closest to 2500\AA\, in the restframe. We then
assume $\alpha$=0.5 for the optical continuum powerlaw slope to derive the
rest-frame, monochromatic optical luminosity at 2500\AA\, in erg 
s$^{-1}$Hz$^{-1}$.  We adopt the most common X-ray/optical measure for
quasars, \aox, defined as the slope of a  hypothetical power-law from
2500\,\AA\, to 2~keV i.e., $\aox\, = 0.3838~{\mathrm log}
(\opteml/\xeml)$.  The mean (median) \aox\, is $1.60\pm0.21$
(1.59) for the binary QSOs, and $1.57\pm0.16$ (1.57) for the
comparison sample.

X-rays in quasars become weaker relative to optical emission as
luminosity increases
\citep{Avni82,Wilkes94,Green95,Steffen06,Green09,Lusso10}. The
\aox\,(\opteml) correlation is persistent across quasar samples, but
has large dispersion.  One possible explanation of the observed trend
is that AGN accretion may transition between accretion states similar
to those of Galactic X-ray binaries (XRBs), where different accretion
rates harden or soften the overall SED.  While AGN vary on much longer
time scales than do XRBs, {\em samples} of AGN may show analogous
trends in SED with luminosity to XRBs \citep{Sobolewska11}.
Statistical tests have shown that the \aox\, correlation is weaker
with redshift, so that the \aox\,(\opteml) relationship is not a
secondary effect of quasar evolution combined with the strong $L-z$
trends of flux-limited quasar samples.  However, some dispute remains
about the influence of selection effects \citep{Yuan98,Tang09}.

\begin{figure*}
% \plotfiddle{PSFILE}{VSIZE}{ROTANG}{HSCALE}{VSCALE}{HRIGHT}{VUP}
%\plotone{opteml_aox0c.eps}
\plotone{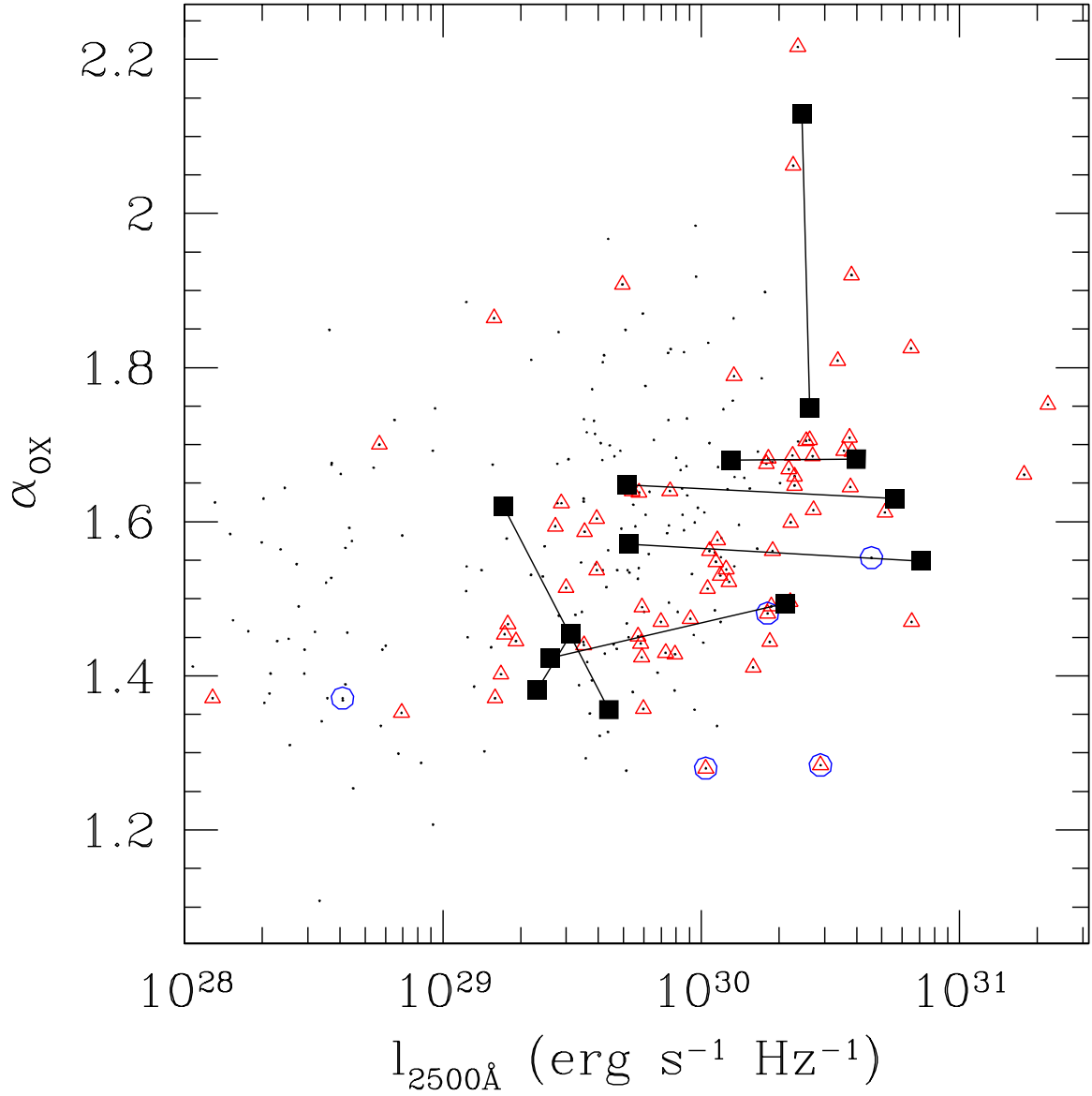}
\caption{\aox\, vs.\ optical 2500\AA\, log luminosity
for the binary QSOs (black squares), with pair members linked
by black lines.  The comparison sample of 264 \Chandra -detected SDSS
QSOs  with $z<1.2$ from \citet{Green09} is also plotted,
for which red triangles indicate spectroscopic redshifts,
and blue circles show radio-loud objects.  
%The best-fit OLS $Y(X)$ regression for the full ChaMP sample is shown
%as a red line with errors. The best-fit relation from
%\citet{Steffen06} is shown as a solid green line.
}    
%\vskip-0.2cm
\label{fig:opteml_aox}
\end{figure*}

If our binary QSO systems genuinely reflect pairs at an unusual
merging stage, perhaps being ignited or exacerbated by an ongoing
merger, we might expect to see differences in the properties of the
AGN involved compared to a random selection of isolated quasars. In
particular, if one or both nuclei in our pairs is being particularly
affected by the merger, we might expect differences in the expected
values of $\aox$, given $\opteml$ for each component of the pair. 
Figure~\ref{fig:opteml_aox} shows \aox\, vs.\ optical
2500\AA\, log luminosity for the binary QSOs (black squares), with
pair members linked by black lines.  The comparison sample of 264
$z<1.2$ SDSS QSOs with \Chandra\ detections from \citep{Green09} is also
plotted, for which red triangles indicate spectroscopic redshifts, and
blue circles show radio-loud objects.  Binary QSO constituents appear
to follow the rather noisy trend of \aox\, with optical luminosity.
Only one QSO, SDSS\,J160602.81+290048.7 falls well away from the
$\aox(\opteml)$ trend, with $\aox =2.2$ at log$\opteml =30.39$.  This
QSO is unusually faint in the X-ray band, and so may be a low redshift
broad absorption line quasar (BALQSO).  This argument is furthered
by the fact that it is probably X-ray absorbed (all 6 measured
photons are above 6\,keV).  The vast majority of
recognized BALQSOs in the SDSS are above redshift 1.6 because only
then is the CIV absorption redwards of the blue cut-off for SDSS 
spectroscopy.\footnote{A much smaller number of the rare low-ionization
  BALQSOs (with BALs just blueward of MgII) are found at lower
  redshifts.}  In most cases, BALQSOs are X-ray weak due to large warm
(ionized) absorbing columns \citep{Green01,Gallagher06}.  BALQSOs
tend to have narrow H$\beta$ broad line components, weak [OIII] lines,
strong optical Fe\,II emission---all of which are
apparent in this object's SDSS spectrum---and be radio quiet.

\begin{figure}
% \plotfiddle{PSFILE}{VSIZE}{ROTANG}{HSCALE}{VSCALE}{HRIGHT}{VUP}
%\plottwo{aoxoptemlADM.eps}{aoxoptemlhistoADM.eps}
\plottwo{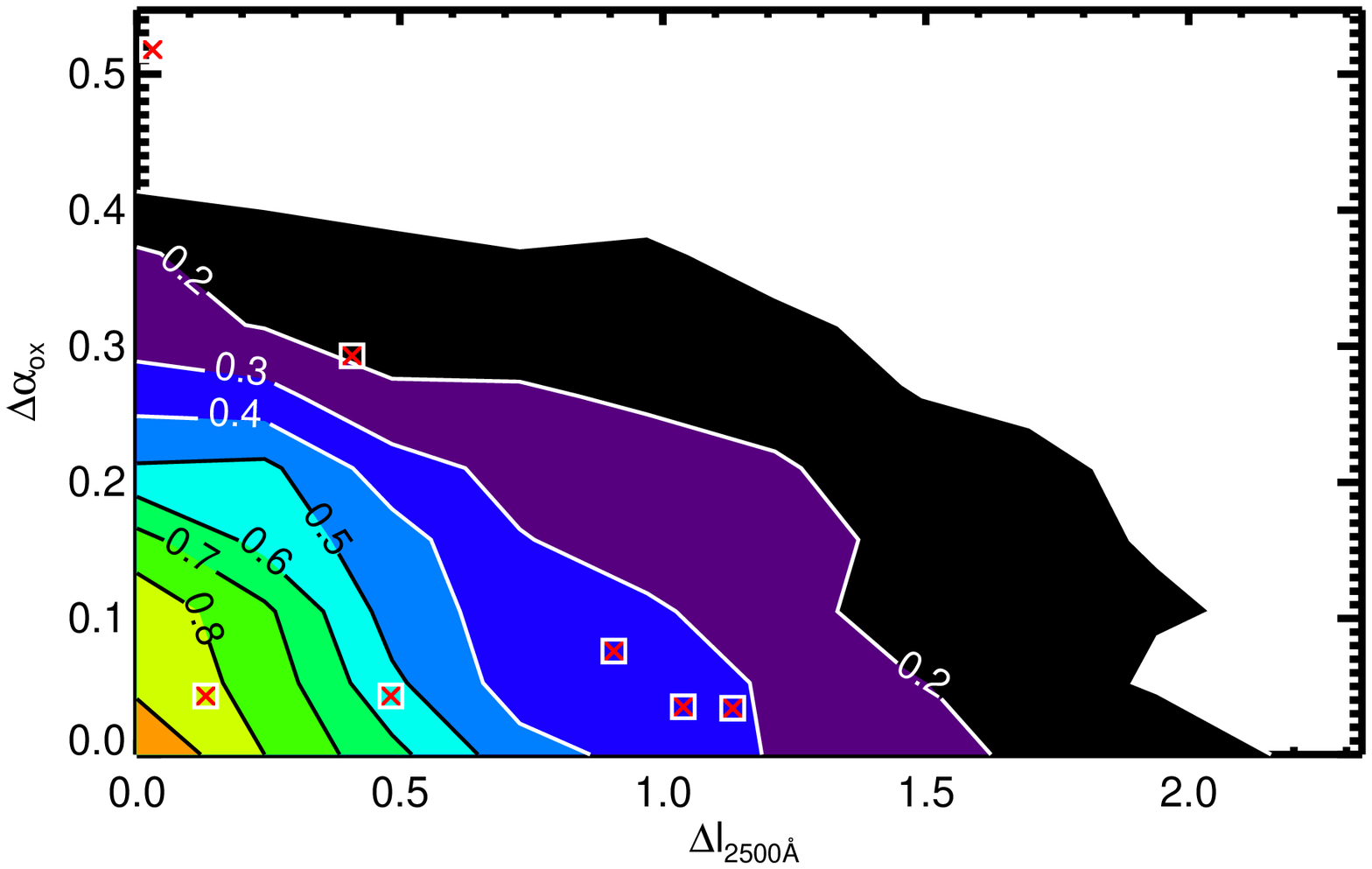}{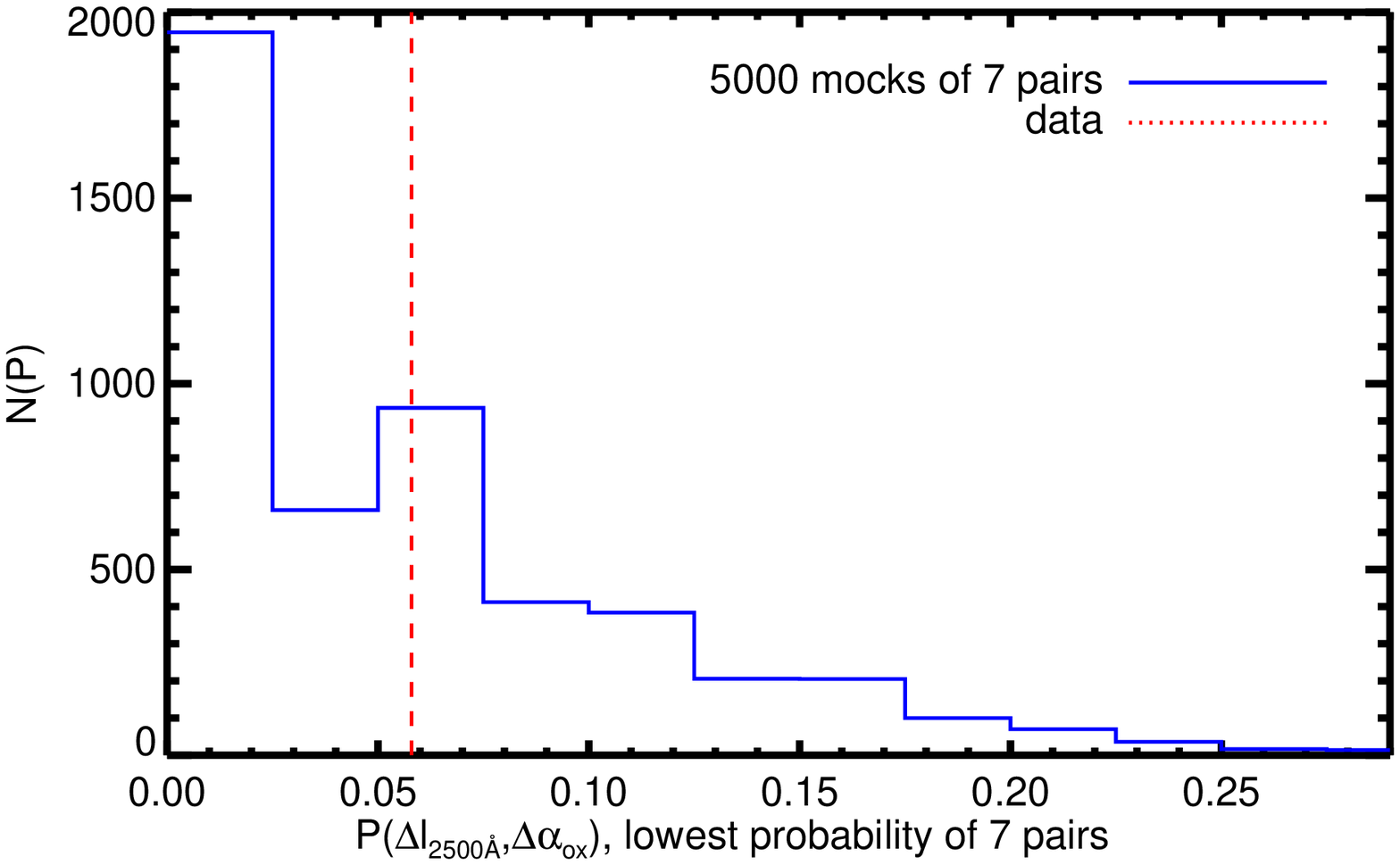}
%\vspace*{-1.25cm}
\caption{LEFT: Distributions of $\Delta\,\aox$,
$\Delta\,\opteml$ for individual quasars and our 7
pairs of close quasars. The contours are the density of 5,000 pairs
of quasars drawn at random from the 264 SDSS quasars in the
ChaMP. Points are the 7 genuine pairs of quasars discussed in this
paper. The most extreme of the 7 data points has a 7\% probability
of being drawn at random from the distribution of possible pairs of
quasars.  RIGHT: As our sample represents drawing 7 pairs of quasars,
rather than just one pair, we repeat our experiment but testing
instead the most improbable pair drawn at random in 5,000 samples of
7 mock pairs. We histogram the value of the contour (i.e. from the
left-hand panel) for the most improbable of the 7 mock pairs. The
dashed line is for the actual binary quasar sample data.}  
%\vskip-0.2cm
\label{fig:ADMfig}
\end{figure}

Criticism of the ensemble trend in \aox(\opteml) observed in samples
of isolated QSOs have been published \citep{Yuan98,Tang09}, charging
that it could be a selection effect caused by the different
dispersions in X-ray and optical luminosity, combined with
flux-limited survey cutoffs.  Binary QSOs represent objects caught at
the same epoch and in the same large-scale environment.  If a
sufficient sample of binary QSOs with dedicated followup also showed
\aox(\opteml) trends, then this might obviate such criticism.
However, since we are testing for potential effects of interaction
between the constituent QSOs, such a test is invalid.  Indeed, we
might test whether the differences between $\aox$ and \opteml\, for
each component of each pair ($\Delta\,\aox$ and $\Delta\,\opteml$,
respectively) are discrepant with the expected trends for isolated
QSOs.  As neither component of our pair is known to be special, we
adopt a one-tailed distribution and only allow these differences to be
positive in value. We establish the background expectation for the
relationship between $\Delta\,\aox$ and $\Delta\,\opteml$ by selecting
5000 pairs at random from the 264 SDSS quasars for which we have
X-ray data from the ChaMP (note that there are then only 
$N\,(N-1)/2$= 34716 possible unique pairs to sample, so our
precision cannot be increased without severely oversampling). In
Figure~\ref{fig:ADMfig} we plot the distribution in density of our
35000 mock pairs in the ($\Delta\,\aox$, $\Delta\,\opteml$) plane
compared to the 7 data points.

It is clear from Figure~\ref{fig:ADMfig} that most of the pairs are
not unusual as compared to background expectation. One of the data
pairs is near the extreme of the distribution, with only 7\% of mock
pairs having similarly extreme values of ($\Delta\,\alpha_{\mathrm{ox}}$,
$\Delta\,\opteml$). However, as we are considering 7
pairs, a result at the 7\% probability level is not unusual---indeed
it should be expected. We demonstrate this further in the right-hand
panel of Figure~\ref{fig:ADMfig}, for which we draw 5000 sets of 7
pairs at random from our 34716 unique mock pairs and plot the contour (from
Figure~\ref{fig:ADMfig}, i.e. the 7\% quoted in this paragraph) of the
least likely pair in the distributions of 7 mock pairs. The histogram
in Figure~\ref{fig:ADMfig} demonstrates that most random sets of 7
pairs have one pair with a probability at the 7\% level. Our data are
therefore not unusual in the ($\Delta\,\alpha_{\mathrm{ox}}$,
$\Delta\,\opteml$) plane, suggesting that either these
values are not unusual for activated nuclei in ongoing mergers, or
that we are not seeing a set of 7 ongoing mergers on the data. 

% xv /data/green/prop/AXAF/wsqps/2008/Spectra/SDSS/spPlot-53496-1578-157.gif

\section{Near Infrared Imaging}
\label{sec:nirim}

To search for extended host galaxy emission and/or morphological
signs of mergers or interactions, we proposed near-IR imaging to
optimize the contrast between the relatively blue quasar point source
emission and the stellar light from the host galaxies. We were
awarded 2 nights to image binary QSOs on Mt Hopkins using the SAO
Wide-field InfraRed Camera (SWIRC) on the 6.5m MMT.  SWIRC has
$2048\times2048$ pixels spanning a $\sim 5.12$\arcmin\, field of view
with 0.15\arcsec\, pixels.  We observed 9 pairs from the larger
binary sample in $J$ (1.2\,$\mu$m)
band on the nights of 25 and 26 March 2010 and obtained between
$6\times90$ sec and $33\times90$ sec dithered images.  We used the
SWIRC pipeline to scale and subtract dark images and remove sky
background from all the images.  The sky image per object frame was
created using SExtractor \citep{Bertin96}. In the
$5.12\arcmin\times5.12\arcmin$ field-of-view, each object frame contained at
least three stars from the Two Micron All Sky Survey (2MASS) point
source catalog \citep{Cutri03,Skrutskie06}, which we
used to calibrate the astrometry of each frame and to determine the
flux zeropoint from the magnitude conversions of \citet{Rudnick01}.
The magnitude where the number counts histogram turns over is a good
general indicator of the limiting magnitude where incompleteness sets
in.  For our shallower field, a typical exposure of 540\,sec results
in a limiting  magnitude of 18.7 while for our deeper fields with
exposure times of 
2970\,sec, the turnover magnitude is 20.3. We then used the ${\tt imwcs}$
software in the WCSTools package \citep{Mink97} to derive sky
coordinates.  We examined the distribution of FWHM for all images
contributing to a given QSO field and excluded any outliers. We then
stacked all the images of a QSO field using the Image Reduction 
and Analysis Facility (IRAF)\footnote{IRAF is distributed
by the National Optical Astronomy Observatory, which is operated by the
Association of Universities for Research in Astronomy, Inc., under the
cooperative agreement with the National Science Foundation.} ${\tt imcombine}$ task---we 
averaged stacked science frames of all astrometrically corrected,
sky-subtracted images, applying a 1$\sigma$-clip. Small
portions of the SWIRC field of view, especially the edges of each
image, were disregarded due to significant contamination from CCD
artifacts. Each of our stacked images contains between 80 and 120
objects consistent with previous $J$-band surveys of the same depth
\citep{Saracco01,Ryan08}.  Seeing at the wavefront
sensor varied from 0.7 to 1.1\arcsec, yielding a typical median FWHM
of 1.25\arcsec on our stacked images.

We obtained SWIRC photometry for a total of four out of the seven
Chandra-observed pairs in our sample. Near-infrared properties are
given in Table~\ref{tnir}. None of our SWIRC QSOs has 2MASS $J$-band 
counterparts, but we find excellent agreement between SWIRC and UKIDSS
for the four QSOs with public UKIDSS photometry.  We detect point
sources for all QSOs, but no evidence of extended emission.
Even SDSS\,J1254+0846, a merger with spectacular tidal tails detected in 
our deep optical imaging \citep{Green10}, shows no SWIRC evidence 
for disturbed morphology.

% Our systems show highly concentrated
% light profiles indicative of a strong concentration of the dust in the
% nuclear region resembling power-law sources rather than interacting
% systems. 
% These might be systems at a later stage
% of their interaction where dust in the extended features has been
% consumed during earlier 
% stages, so the tidal tails are dominated by young stars similar to
% the example of NGC2623 (Scoville et al. 2000; Rossa et al. 2007). Our
% QSOs appear morphologically similar in the near-infrared to PG QSOs
% (Veilleux et al. 2009) consistent with our comparison with single
% quasars presented in \S~=\ref{sec:compare}.

\section{Quasar Spectral Energy Distributions}
\label{sec:SEDs}

To characterize SEDs, estimate bolometric luminosities and check
for starburst activity we fit template SEDs to all QSO pairs in our sample.
We fit to all of  the near-infrared, optical and X-ray fluxes described
above, using a library of 12 templates: a radio-quiet
Type~1 quasar, a luminosity-dependent radio-quiet Type~1 quasar
(i.e., where \aox\, follows the known trend with \opteml), two
Type~2 (narrow line) Seyferts, four starburst and four composite templates
\citep{Ruiz10}.  We fitted all SEDs by using the $\chi$$^{2}$
minimization technique of \citet{Ruiz07,Ruiz10}. Existing optical
spectroscopy for each of our sources removes the uncertainty of using
photometric redshifts for the SED fitting and provides a direct
testbed for the accuracy of the fitting method.  Broad-band SEDs for
all 14 sources are given in Figure~\ref{fig:seds1}. Table~\ref{tseds} 
gives the different parameters of our best SED fits. 

In all 14 cases, either a radio quiet Type~1 QSO (\citealt{Richards06}
for $\nu>10^{12}$ Hz and \citealt{Elvis94} for $\nu < 10^{12}$
Hz) or an AGN luminosity dependent template \citep{Hopkins07a} is
needed to fit the photometry, consistent with the fact that each of our
sources are spectroscopically identified as broad line QSOs. Nonetheless,
for at least one component in 5 out of 7 pairs---8 out of 14
sources in total---the best-fit SED requires an additional starburst
component.  

We can contrast this fraction (8/14, or 57\%) with that for single
broad-line AGN (BLAGN) with spectroscopy from the ChaMP
\citep{Green04,Green09}.  Of 758
spectroscopically identified ChaMP BLAGN, 184 have near-IR photometry, and
98 of those (54\%) require a template with a starburst component.  If we
further restrict the ChaMP sample to the same $0.44 < z < 1$ redshift
range as our binary QSOs, the fraction does not change (53\%).  To
summarize, our binary QSO SEDs are no different than those of
X-ray-selected QSOs with optical and near-IR counterparts in the
ChaMP. 

When the AGN and/or starburst component contribution is estimated over
the 10$^{14}$ - 10$^{18}$ Hz wavelength range, where we have available
photometry, the luminosity of all 14 sources appears to originate
mainly from an AGN component ($>$55$\%$). However, in the case of the
8 sources that require a starburst component, star-formation
activity contributes at least 20$\%$ of the luminosity emitted between
10$^{14}$ -- 10$^{18}$ Hz.  When we integrate luminosity 
over  the entire radio to X-ray wavelength range, starburst activity
becomes the dominant component ($>$90$\%$) in 6 cases, which may be
indicative of intense star-formation events in their hosts.  
We warn, however, that the bulk of the starburst template contribution
comes from longer wavelengths---far-IR to radio---than our available data.

To seek further constraints on the mid- to far-infrared spectral
regions, we have cross-correlated our binary QSO sample with the 
Preliminary Data Release catalog (April 14, 2011) of the 
Wide-field Infrared Survey Explorer (WISE; \citealt{Wright10}).
The catalog covers about 23,600\,\degsq\, with typical stacked
exposures near 100\,sec for at 3.4, 4.6, 12, and 22\,$\mu$m
with an angular resolution of 6.1\arcsec,  6.4\arcsec, 6.5\arcsec\,
and 12.0\arcsec, respectively.  Although WISE is thus unable to
resolve the two  components of our pairs, our systems are bright
enough for WISE to be able to detect the emission from the paired
system.  Four of our pairs (J0740+2926, J0813+5416, J1508+3328,
J1606+2900) have WISE counterparts with detections in all four WISE
bands. 
% ranging from 13$<$3.4$\mu$m$<$14.109, 11.791$<$4.6$\mu$m$<$12.784,
%9.218$<$12$\mu$m$<$9.931, 6.769$<$22$\mu$m$<$7.52 Vega magnitudes. 
To test our SED fitting results at these longer wavelengths, we have
utilized the WISE detections to create the combined "pair" SEDs of the
four pairs with mid-infrared detections. For each pair, we have simply
added the near-infrared, optical and X-ray fluxes together 
for the two binary QSO components and appended the WISE fluxes into
the total pair SED. We then fit the summed SEDs for each pair. In
the case of J0740+2926 and J1508+3328, the best solution is a pure
AGN. In the case of J0813+5416 and J1606+2900, the best solution is a 
composite SED with the AGN contributing up to 80$\%$  and 87$\%$ of
the emission, respectively with the remainder coming from a starburst
component.  The agreement between these results from the individual and
paired fitting is almost perfect.  Disagreement for the case of
J1508+3328, with no starburst component required in the summed
SED with WISE, is likely due to the fact that it has the smallest 
total starburst contribution of all summed pairs ($\lax 12$\%;
see Table\,\ref{tseds}).

The predicted SED fits suggest that several of the QSO pairs in
our sample may have significant ongoing starformation, detectable
even in the presence of luminous QSO emission.
Interestingly, the one system, SDSS J1254+0846 \citep{Green10}, that
is known to inhabit a merger, does not require a significant
contribution from star formation.  Previous studies of X-ray-selected
\citep{Trichas09} and spectroscopically confirmed \citep{Lutz08, Trichas10,
  Kalfountzou10, Symeonidis10} QSOs with far-infrared
detections have shown that the vast majority of these sources are
composite objects with very strong ongoing starburst events.  While
these studies were focused on the brightest and rarest examples,
subsequent studies of submillimeter-detected Type~1 QSOs
\citep{Lutz10,Hatziminaoglou10} have made it clear that the
sub-millimeter colors of Type~1 QSOs are similar to those of
star-forming galaxies. This hints at an emerging picture where
star formation is present in the environs of all AGN, consistent with merger
models like those of \citet{Hopkins05}. On the other hand, in the
local Universe, all black hole accretion as detected by hard X-rays is
strongly \emph{disassociated} with star formation implying that there
is a fundamental 
anti-correlation between the two that is not a selection effect
\citep{Schawinski09}. In the latter case, the prediction of starburst
activity in the majority of the QSO pairs in our sample has strong
implications for the dynamics of these merger systems that should be
further investigated.

\section{Search for Evidence of Host Clusters}
\label{sec:clusters}

\subsection{X-rays from Host Clusters}
\label{sec:cxocluster}

Despite the presence of bright quasar emission, we know
that Chandra can detect extended cluster emission against a typical
ACIS-S background with $\sim$50 diffuse counts or more in any of our
fields \citep{Green02, Aldcroft03}.  To ensure that we could
detect clusters  as weak as $\sim 0.1\, L_X^*$, we slightly increased
our proposed Chandra exposure times above what was required for the
QSOs themselves (see \S\,\ref{sec:cxo} where necessary, extrapolating
from typical cluster relationships ($L_{0.5-2\,keV} \geqsim 3\times
10^{43}$; \citealt{Mullis04}).  For two pairs we thus increased 
exposure times slightly: SDSSJ0740 (+5\,ksec) and SDSSJ1606
(+6\,ksec). 

The ACIS image of SDSS\,J1158+1235 (obsid 10314)
displays significant extended X-ray emission---but, it is
43\arcsec\, SSW of the QSO pair's midpoint. The peak of the extended
X-ray emission is coincident with  a luminous $i=17.18$ absorption-line 
galaxy at $z=0.2652$ (SDSS\,J115821.96+123438.6).  %  588017566561206349
At absolute magnitude $M_i\sim -23.71$, this is clearly the cD galaxy
of an X-ray cluster. Using a circular aperture of 24\arcsec\, radius
for the cluster,  and a background annulus from 62 -- 110\arcsec\,
excluding  all detected source regions, we derive $301\pm19$ counts
from the cluster.  Assuming a Raymond-Smith plasma with $T=2$\,keV and
metallicity 0.2 solar, we derive 
using the Chandra Portable, Interactive Multi-Mission Simulator
PIMMS\footnote{http://cxc.harvard.edu/toolkit/pimms.jsp,
originally \citet{Mukai93}.} $f(0.5$--$2\,\mathrm{keV})=3.97\times  
10^{-14}$\,\fcgs, and a luminosity of $8.2\times 10^{42}$\,\lcgs.
Errors on these values are dominated by the spectral assumptions, and
are likely to be $\sim$15\%.

Otherwise, no significant extended emission sources are evident to the
eye on the ACIS-S3 images in the immediate neighborhood of the QSO
pairs.  When searching for faint extended sources, however, it is
important to minimize background contamination.  The ACIS particle
background increases significantly below 0.5\,keV and again at high
energies.  To optimize detection and visual inspection of possible
weak cluster emission, we first filtered the cleaned image to include
only photons between 0.5 and 2\,keV.  Around the QSO positions as
detected by {\tt wavdetect}, we masked out pixels within twice the
radius that encompasses 95\% of the encircled energy. (The 95\% PSF
radius at 1.5\,keV is about 2.06\arcsec.)  For visual inspection, we
also excised regions around all other detected sources, corresponding
to 4 times the 4$\sigma$ Gaussian source region output of {\tt
  wavdetect}.

We then generated 7 annuli of 50\,kpc projected width each,
starting at $R=75$\,kpc from the mean of the detected QSO
coordinates.\footnote{Since SDSS\,J1606+2900B was not detected
by \Chandra, we simply use its optical position from SDSS.}
Though the sample redshifts range from 0.44 to 0.978, these radii only
differ slightly between the targets (dispersion in the mean is about
13\%), so we used a single set of six 7\arcsec\ annuli from 10
to 52\arcsec.  We set a background annulus  from 60 -- 110\arcsec,
and calculated radial surface-brightness profiles.

There are just 2 fields with radial profiles that rise consistently
inward towards the QSOs.  For SDSS\,J0740+2926 (obsid 10312), the profile
arises from some faint diffuse emission with a centroid about
7.5\arcsec\ NW of the mean QSO positions.
% 115.05368 29.44749 = 07 40 12.88 +29 26 50.96
% ellipse(4095.5,4074.4998,10.5,26,15.513012)
% see 10312mosaic.jpg
The emission only encompasses about $9.8\pm3.6$ net (0.5-2\,keV) counts,
and there is at least one other such source in the field, so we
discount its reality. 
% fx(0.5-2)=  1.674E-15
%   z      DcMpc   VGpc   kpc/as distmod  DlMpc  logSphere
% 0.9780  3159.43  132.103  7.74  43.979  6249.3  57.6696
%  log Lx = 42.893
The other field with a suggestive radial profile is that of 
SDSS\,J1508+3328 (obsid 10317), which similarly shows an apparent 
weak diffuse emission region at 6.8\arcsec\ W of
the mean QSO position, with $14\pm4$ net (0.5--2\,keV) counts.
There is no other such source apparent in the field.
%  fx(0.5-2)= 1.578E-15
%  DcMpc   VGpc   kpc/as distmod  DlMpc  logSphere
% 2912.79  103.518  7.52  43.690  5470.2  57.5539
%  log Lx = 42.75
These weak excesses may represent the emission from nearby galaxies
that fall individually below the detection level, or from a weak ICM

The ACIS image of SDSS\,J0813+5416 (obsid 10313) shows signs in the
smoothed image of extended emission that could be more filamentary in
shape, and so would not register as a significant trend in a radial
profile plot. The emission appears to extend about 80\arcsec\, from
SE to NW. Excluding the QSO regions, and using elliptical source and
background apertures (of about  0.7 sq.\ arcmin area), we tally
$64\pm12$ net source counts.  There are no obvious optical
counterparts that might be galaxies associated with a cluster
merger or cosmological filament.  Assuming a Raymond-Smith plasma with  
$T=2$\,keV and metallicity 0.2 solar, we derive
$f(0.5$--$2\,\mathrm{keV})=7.43 \times 10^{-15}$\,\fcgs. If at the $z=0.779$
redshift of the QSOs, the cluster luminosity is $2.0\times 10^{43}$\,\lcgs. 
% Cycle 10 rate to 0.-2keV flux conversion factor 3.548e-12 erg cm-2 
%   z      DcMpc   VGpc   kpc/as distmod  DlMpc  logSphere
% 0.7790  2653.84  78.291  7.23  43.370  4721.2  57.4260
% calc -14.1292+57.4260 = 43.2968

\subsection{Optical Imaging}
\label{sec:optim}

\subsubsection{Kitt Peak}
\label{sec:optim4m}

To study each component of each binary quasar in the optical, and to search for further signs
of merger activity or local galaxy overdensities, we 
imaged six binary QSOs at Kitt Peak National 
Observatory (KPNO) using the 4m Mayall telescope on the nights of 2009
March 17--19.  All images were acquired with the MOSAIC 8K camera
($8192\times 8192$ pixels; $0.26\arcsec~\mbox{pixel}^{-1}$) in one or
more filters using the $r$, $i$, and $z$ bandpasses.
Integration times ranged from 900 to 9000 s, depending on the filter
and the redshift of the target binary quasar. The seeing varied during
the observing run from 0.77{\arcsec} to 1.69{\arcsec} (FWHM), as
measured from the combined frames.

Image reduction was conducted using the {\tt mscred} package within IRAF.
Processing of the raw images involved the standard procedure of bias
correction and flat-fielding using dome flats and deep sky exposures. After
initial processing, individual images were astrometrically corrected and
median combined to yield a higher S/N image.

Object detection and photometry was conducted using SExtractor
\citep{Bertin96} via the ChaMP image reduction pipeline \citep{Green04}.
Since all images were acquired during non-photometric sky conditions,
instrumental magnitudes were transformed to the standard system by
calibrating to overlapping SDSS DR7 data using dereddened
magnitudes.\footnote{We compare SDSS {\tt model\_Mag} to SExtractor
  MAG\_AUTO values.}  We typically achieved magnitude
limits\footnote{We quote the magnitude where the number counts
peak in a differential (0.25\,mag bin) number counts histogram. This
corresponds approximately to 90\% completeness in the magnitude 
range 20--25, and is typically about 1\,mag brighter than the
$5\sigma$ limiting magnitude \citep{Green04}.} of 24--24.5 in $i$
and 24 in $r$. We examined all images for any evidence of extended
emission or disturbed morphology. This led to the discovery of
tidal tails in both $r$ and $i$ band images of SDSS\,J1254+0846.
However, due to poor weather, we were only able to  image 
6 of the 7 fields in this \Chandra\ subsample (all but 
SDSS\,J1418+24410), and obtained imaging in more than one band 
for only 2 fields: around SDSS\,J1158+1235 and SDSS\,J1254+0846.
This precluded an effective photometric search for galaxy overdensities, 
described below.

\subsubsection{SDSS}
\label{sec:optimsdss}

With optical imaging of adequate depth, we can photometrically detect an overdensity
of galaxies---because early-type galaxies at a given
redshift have a narrow range of colors which form a cluster
``red-sequence'' \citep{Gladders00} in their color-magnitude
diagram (CMD).  In the neighborhood of a QSO pair, the most convincing
optical cluster detection would have a large number of optical
galaxies clustering at small distances from the QSO pair center, and
those galaxies would have well-measured colors clustering at small
distances from a single locus in the CMD.  We therefore define a
distance- and error-weighted color mean (DWCM), given by

\begin{equation}
\mathrm{DWCM}=\sum\frac{(r-i)_{j}}{\sigma^{2}_{j}} \Big/ \sum\frac{1}{\sigma^{2}_{j}} \,\, ,
\end{equation}

\noindent where $\sigma^{2}_{j}=\sigma^{2}_{(r-i)_{j}}+R^{2}_{j}$, and $R$ is
the projected distance from the mean quasar pair position in units of Mpc
at the QSO redshift. Thus a bright galaxy with a small color error could
contribute as much to the DWCM as a fainter galaxy closer to the
center point.  With the 1\,Mpc scaling, a 250\,kpc projected 
galaxy distance and a typical color error of 0.25 contribute about
equally to the weighting.   Using the same DWCM  
calculation for any number of randomly-chosen locations in the same
large-field optical image of our quasar field allows us to quantify
the significance of the DWCM measured around our QSO pairs, 
in a way that naturally accounts for the characteristics of the
relevant imaging such as depth and image FWHM.  

We selected all objects within 48\arcmin\, of each quasar pair's mean
position from the SDSS DR8 using the CasJobs query interface.  The
turnover (model) magnitudes are $r\sim22.8$, $r\sim22.2$ and
$i\sim21.4$, which correspond to about 50\% completeness for point
sources\footnote{Based on comparison to
  http://www.sdss.org/dr7/products/general/completeness.html}.  We
included only objects with $15<r<22.5$, for which the median error in
$(r-i)$ and $(i-z)$ color are 0.105 and 0.192, respectively.  We
calculated the DWCM in both $(r-i)$ and $(i-z)$  for 1000 random
positions for each of our quasar pairs, and at the actual
position of the quasar pair, always in an annulus between 25\,kpc and
500\,kpc projected radius at the QSO redshift.
We then compare the DWCM to the expected observed-frame
color of the galaxy red-sequence for the Schechter magnitude
$M^{\ast}$ based on the redshift of each quasar pair, adopting the red
sequence models of \citet{Kodama97} transformed to the SDSS filters
(T. Kodama 2004, private communications). 

If the DWCM at the QSO position is appropriate for the red-sequence
color expected at the pair redshift, and the red-sequence scatter is
small, this could be especially convincing evidence for a physical
galaxy cluster. To estimate the prominence of the red-sequence, we
calculate the variance in DWCM and compare the results for our set of
random locations with results for the location of the quasar pair. 

%   z   appmag_r appmag_i   rmi  rmislope  rmiint  imz   imzslope  imzint
%0.9750 23.8582  22.7554  1.1028 -0.0201  1.5611  1.0519 -0.0420  2.0084 
%0.7750 22.9963  21.7459  1.2504 -0.0368  2.0496  0.7576 -0.0265  1.3338 
%0.6000 21.9643  20.8003  1.1640 -0.0436  2.0714  0.5168 -0.0151  0.8309 
%0.4375 20.7673  19.9857  0.7816 -0.0234  1.2485  0.4404 -0.0129  0.6989 
%0.5750 21.8169  20.6902  1.1267 -0.0419  1.9939  0.5014 -0.0151  0.8136 
%0.8750 23.3986  22.2703  1.1283 -0.0200  1.5737  0.9848 -0.0409  1.8950 
%0.7750 22.9283  21.6753  1.2530 -0.0368  2.0496  0.7595 -0.0265  1.3338 

% See /data/green/prop/AXAF/wsqps/2008/Paper/DWCM/2011apr13

\begin{figure*}[t]
% \plotfiddle{PSFILE}{VSIZE}{ROTANG}{HSCALE}{VSCALE}{HRIGHT}{VUP}
%\plotfiddle{1158imz.eps}{3in}{-90}{200}{200}{0}{0}
%\plotfiddle{1158rmi.eps}{3in}{-90}{200}{200}{200}{0}
%\vspace{-5in}
%\plotone{1158dwcm.eps}
\plotone{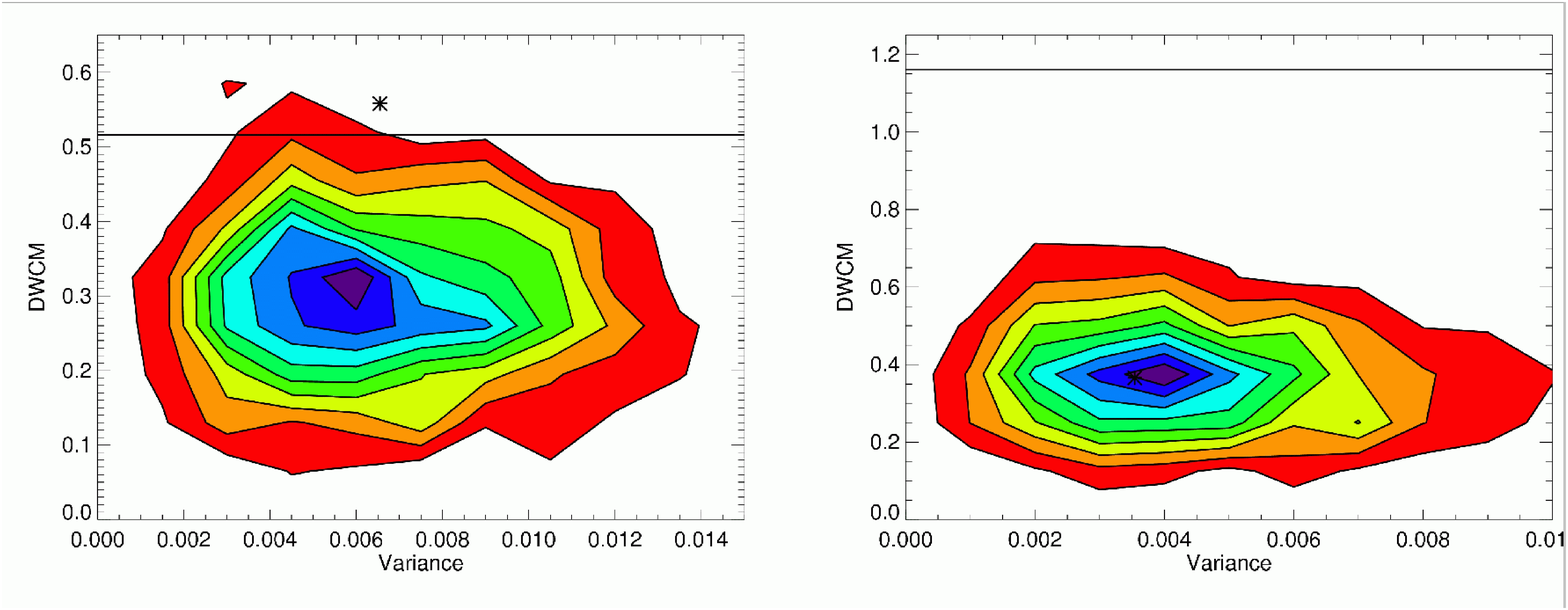}
\caption{Test of optical cluster significance.
(LEFT) Contours from 10 to 90\% in relative frequency show the
distance-weighted ($i-z$) color measure (DWCM) vs.\ its variance for
1,000 random positions within the (2\,deg) SDSS field of the QSO
pair SDSS\,J1158+1235, calculated using the DWCM prescription
described in \S\,\ref{sec:optimsdss}.  The asterisk marks the DWCM 
and its variance calculated at the actual position of the quasar pair
SDSS\,J1158+1235.  The expected location  ($i-z=0.98$) of the galaxy red
sequence at $m^{\ast}$ based on the mean redshift  $z\sim0.878$ of the 
quasar pair is shown as a solid horizontal line, adopting the red
sequence models of \citet{Kodama97}. While the variance is typical
($\sim\,80\%$) for the field, the actual mean color (DWCM) is both
unusual and close to the expected red-sequence color.  (RIGHT) The
DWCM vs.\ its variance for $(r-i)$ shows that the expected
red-sequence value at the midpoint of the QSOs' positions (asterisk) is 
entirely typical for random positions in the field, and far from the  
expected red-sequence color (horizontal line).}
\label{fig:dwcm}
\end{figure*}

Only the SDSS\,J1158+1235 position shows a DWCM both close to the
$(i-z)$ color expected for an overdensity and significantly different from 
contours for a DWCM derived by sampling random positions in the field.
However, this test does not hold for $r-i$. Both plots are shown in
Figure~\ref{fig:dwcm}.

We conclude that we detect no significant galaxy density enhancements
of the color and magnitudes expected for early-type galaxies at the
redshift of our QSO pairs.  However, we note that across
the full redshift range of our sample, the expected apparent Schechter
magnitude in the $r$ band ranges from 20.8 at $z=0.44$ to 23.9 at
$z\sim 1$.  Therefore, for objects beyond $z\sim0.7$, we detect few
bright galaxies of the expected color. 

\section{Discussion}
\label{sec:discuss}

% Why we studied these guys, what we found, and what it means.

In binary quasars, both host galaxies would already have substantial
SMBH and pre-existing stellar bulges, so must have a significant
history of accretion before the observed episode of simultaneous 
activity.  But at any moment in time (i.e. when observed) either or both 
component QSOs might otherwise be quiescent.  Our intent in this work is to study
both the SEDs and environments of binary quasars to search for signs
that interaction might indeed be triggering the currently observed
activity.  One alternative to the interaction/triggering 
interpretation is simply that QSOs are more likely to be found
in overdense regions, with a QSO pair likely to be found in some
fraction of those, perhaps more likely in those inhabiting the largest 
overdensities.  

To probe for host signatures of merging or triggering is challenging
in  luminous QSO pairs, both because of their bright nuclei, and
because they are found at significant redshifts, making host imaging
difficult. Nevertheless, in our small \Chandra\ sample of 7 binary
quasar pairs, we have discovered one clear example of an interacting
system in our lowest-redshift pair, SDSS\,J1254+0846 \citep{Green10}. 
In this paper we pursued two other potential indicators of unusual accretion or
star-formation activity: multiwavelength spectral energy distributions
and environment.  

\subsection{Spectral Energy Distributions}
\label{sec:discussSEDs}

Analyzing results from published optical/infrared photometry and our
own \Chandra\ observations, we find that the SEDs are consistent with
those of isolated QSOs.  Their X-ray spectra are typical, and show no
sign of excess absorption that might be expected in systems with
accretion rates enhanced by interactions that dissipate angular
momentum of gas.  The ratio of optical to X-ray emission in these
QSOs, characterized here by \aox\, is also typical, both in its
distribution and in its correlation with luminosity.  Such a finding
might be expected based on these pairs' original selection by their
typical optical quasar colors \citet{Myers08}.

Based on our SED fits, the available optical and near-IR SEDs show
possible evidence for enhanced star-formation activity, because the
best fit requires a starburst template in addition to a standard QSO
template.   
% EVIDENCE for STBST CONTRIB TO SEDs
It would be of great interest to test more robustly whether this
tendency is statistically different from isolated QSOs. In a
subsequent paper (Trichas et al. 2011), we are planning to utilize the
large number of ChaMP spectroscopically identified isolated QSOs to
compare their SEDs to a much larger sample of spectroscopically
identified QSO pairs (e.g., \citealt{Myers08}). Inclusion of {\em WISE}
and/or {\em  Spitzer}, {\em Herschel} and {\em ALMA} photometry would
greatly improve over current constraints on star-formation activity.

% FUTURE LOOK AT DYNAMICAL vs SED CORRELATIONS
For a larger binary quasar sample, we can also correlate the SED
characteristics and $L_{\rm Bol}$ with  dynamical characteristics like
$R_p$ and $\Delta v$---do smaller separations and/or lower velocities
result in more luminous, high column systems? 

Today's favored models (e.g., \citealt{Hopkins08}) associate luminous
AGN activity with major mergers, so the lack of significant SED-based
evidence for interactions is interesting.  Proximity does not dictate
merging.  Differences between predicted galaxy-galaxy merger rates can
be significant (factor $\sim 5$; \citealt{Hopkins10}), attributable at
least in part to the treatment of the baryonic physics, especially
those in satellite galaxies.

Over a redshift range similar to our sample, \citet{Cisternas11} find
no significant difference in the fraction of (HST ACS) distorted
morphologies between X-ray active and inactive galaxies in the COSMOS
field.  Consequently, they argue that the bulk of black hole 
accretion has not been triggered by major galaxy mergers, but more
likely by alternative mechanisms such as internal secular processes or
minor interactions.

\subsection{Environments}
\label{sec:environments}

We find no evidence that these pairs inhabit significant
galaxy overdensities based on a search for red-sequence galaxies
in SDSS optical imaging.  Neither do they show signs of 
inhabiting a hot ICM that might be associated with a significant
cluster of galaxies or a massive dark matter halo.

While we might hope that quasars---especially binary quasars---would
be signposts for high-redshift clusters, this has not turned out to be
the case.  At low redshift, there are just a handful of X-ray
clusters associated with quasars or powerful radio galaxies at lower
redshifts (e.g., Cygnus A: 3C 295, \citealt{Allen01}; IRAS 09104+4109,
\citealt{Iwasawa01}; HS1821+643, \citealt{Russell10}), and even fewer
at high redshifts \citep{Siemiginowska10}.

The fraction of galaxies hosting AGN evolves with cosmic time
\citep{Shi08,Martini09,Shen10a,Haggard10} and is likewise affected by
environment (e.g., \citealt{Strand08}).  Luminous quasars and intense
star formation activity both tend to be found at $z\,\geqsim\,1$, where
there are still very few massive clusters known.  While the space
density of luminous AGN decreases drastically toward the present day
(e.g., \citealt{Silverman08}), the clusters are assembling. Burgeoning
detections of galaxy clusters based on the Sunyaev-Zeldovich effect
(SZE) \citep{Staniszewski09, Vanderlinde10} may help widen the overlap.

Whether AGN favor or eschew cluster environments in a given
epoch is another question.  At low redshifts, the fraction of galaxies
that host X-ray AGN  appears to be the same in clusters as in the
field \citep{Haggard10}, although the fraction in clusters may evolve
more rapidly than the field \citep{Martini09}.  

For galaxies at low redshift ($z<0.1$), lower density environments
have fractionally more galaxy pairs with small projected separations
and relative velocities \citep{Ellison10}.  Conversely, selection of
pairs by small projected separation and low $\Delta v$ tends to
select lower density environments, an effect which may apply here as
well, since we restricted the parent binary quasar sample
to velocity differences $\Delta v < 800$\,km\,s$^{-1}$ and separations
$R_{p}<30$\,kpc.  Galaxies in the lowest density environments  show
the largest star formation rates and asymmetries for the smallest
separations, suggesting that triggered star formation is seen only in
lower density environments \citep{Ellison10}.  Whether this does or does not
apply to AGN triggering remains unclear.

%% Here is the endmatter stuff: Supplementary Info, etc.
%% Use \item's to separate, default label is "Acknowledgements"

\acknowledgments
 Support for this work was provided by the National Aeronautics and
Space Administration  through \Chandra\, Award Numbers GO9-0114A
and GO9-0114B issued by the \Chandra\, X-ray Observatory Center, which 
is operated by the Smithsonian Astrophysical Observatory for and on 
behalf of the National Aeronautics Space Administration under contract 
NAS8-03060. Discovery optical images were obtained at Kitt Peak National
Observatory, National Optical Astronomy Observatory, which is operated
by the Association of Universities for Research in Astronomy (AURA)
under cooperative agreement with the National Science Foundation.  
This research has made use of data obtained from the \Chandra\ Data
Archive and software provided by the \Chandra\ X-ray Center (CXC) in the
application packages CIAO, and Sherpa.  This paper has used data from
the SDSS archive. Funding for the SDSS and SDSS-II has been provided
by the Alfred P. Sloan Foundation, the Participating Institutions,
the NSF, the US Department of Energy, the National Aeronautics and Space
Administration (NASA), the Japanese Monbukagakusho, the Max Planck
Society, and the Higher Education Funding Council for England. The
SDSS website is at http://www.sdss.org/. 

Thanks to Doug Mink and Bill Wyatt (SAO) for help with the SWIRC
pipeline.

{\it Facilities:} 
\facility{Mayall ()}
\facility{Magellan:Baade ()}
\facility{CXO ()}, 
\dataset[ADS/Sa.CXO#obs/10312]{\Chandra\ ObsId 10312}
\dataset[ADS/Sa.CXO#obs/10313]{\Chandra\ ObsId 10313}
\dataset[ADS/Sa.CXO#obs/10314]{\Chandra\ ObsId 10314}
\dataset[ADS/Sa.CXO#obs/10315]{\Chandra\ ObsId 10315}
\dataset[ADS/Sa.CXO#obs/10316]{\Chandra\ ObsId 10316}
\dataset[ADS/Sa.CXO#obs/10317]{\Chandra\ ObsId 10317}
\dataset[ADS/Sa.CXO#obs/10318]{\Chandra\ ObsId 10318}

\appendix
\section*{Appendix: Explicit X-ray Flux and Luminosity Calculations}
\label{sec:xcalcs}

We often assume the monochromatic flux density for an underlying
intrinsic power-law to have form $f\propto\nu^{\alpha}$, where
$f$ is the monochromatic flux (e.g., in \fcgs Hz\mone) and $\nu$ is
the power-law frequency index.  For X-rays, the photon index $\Gamma$
is more commonly used, where $\alpha=(1-\Gamma)$.   

We fit an X-ray power-law spectral model  
 $$ N(E) = N(E_0)\, 
\Bigl(\frac{E}{E_0}\Bigr)^{(1-\Gamma)}\, \exp [-N_H^{Gal}\sigma(E)
              - N_H^z\sigma(E(1+z)) ] $$ 
to the X-ray counts as a function of energy, where $N(E_0)$ is the   
normalization in photons cm$^{-2}$ sec$^{-1}$ keV$^{-1}$ at a chosen
reference energy $E_0$, $\Gamma$ is the photon index,
and $\sigma(E)$ is the absorption cross-section.  We fix $N_H^{Gal}$
at the  appropriate Galactic neutral absorbing column, and allow
for an intrinsic absorber with neutral column $N_H^z$ at the source
redshift. 

The X-ray monochromatic energy flux
without the effects of absorption is   
   $$ f(E) = E\,N(E) = E\,N(E_0)\,\,\Bigl(\frac{E}{E_0}\Bigr)^{(1-\Gamma)}$$
in keV cm$^{-−2}$ sec$^{-−1}$ keV$^{−-1}$.  Then, since
   $ f(E_0) = E_0\,N(E_0),$
we can express the  monochromatic energy flux as
   $$ f(E) = f(E_0)\,\Bigl(\frac{E}{E_0}\Bigr)^{(1-\Gamma)}.$$
To obtain the more standard units of erg cm$^{-2}$ sec$^{−1}$ Hz$^{-1}$,
%    $$ f(\nu) = f(\nu_0)\,\Bigl(\frac{\nu}{\nu_0}\Bigr)^{(1-\Gamma)},$$
multiply by $
6.629\times 10^{-27}$ (from conversion factors $1.602\times
10^{-9}$\,erg/keV and Hz$^{-1}$=$h$\,keV$^{-1}$ where  $h =
4.138\times 10^{-18}$\,keV/Hz).  

The integrated flux observed between energies $E_1$ and $E_2$ is
    $$ F = \int_{E_1}^{E_2}\,f(E)\,dE  \\
    = \frac{f(E_0)}{E_0^{(1-\Gamma)}}\,\frac{[E_2^{(2-\Gamma)} -
        E_1^{(2-\Gamma)}]}{(2-\Gamma)} 
    = \frac{N(E_0)}{E_0^{-\Gamma}}\,
    \frac{[E_2^{(2-\Gamma)} - E_1^{(2-\Gamma)}]}{(2-\Gamma)}. $$
If $F$ above is in units of keV cm$^{-−2}$ sec$^{-−1}$, multiplying by
1.602$\times 10^{-9}$ yields observed broadband flux in erg cm$^{-2}$
sec$^{−1}$. 

Note that as $\Gamma\rightarrow 2$, via L'Hospital's rule
$F\rightarrow  \frac{N(E_0)}{E_0^{-\Gamma}}\,\,ln(E_2/E_1)$.  
Note also that to     convert from  one broadband flux (or luminosity)
to another 
   $$ \frac{F(E_3-E_4)}{F(E_1-E_2)}= \frac{[E_4^{(2-\Gamma)} -
        E_3^{(2-\Gamma)}]}{[E_2^{(2-\Gamma)} - E_1^{(2-\Gamma)}]}.$$ 

%For  $f\propto\nu^{\alpha}$, the rest-frame
%spectral (monochromatic) luminosity $l$ at frequency $\nu_1$ is related
%to the monochromatic flux at $\nu_2$ via
%$$l(\nu_1) = \frac{ 4\,\pi\,d_L^2 }{ (1+z)^{(1+\alpha)} } 
%    \Bigl(  \frac{\nu_1}{\nu_2} \Bigr)^\alpha f(\nu_2)$$ 
%or
%$$l(E) = \frac{ 4\,\pi\,d_L^2 }{ (1+z)^{(2-\Gamma)} } 
%    \Bigl(  \frac{E}{E_0} \Bigr)^{(1-\Gamma)} f(E_0)
% =  4\,\pi\,d_L^2  (1+z)^{(\Gamma-2)}  f(E)$$ 

Due to the redshift, the measured spectral flux $f_{\nu}$ is
related to the spectral rest-frame luminosity $L_{\nu^{\prime}}$, where 
$\nu^{\prime}= (1+z)\,\nu$, as
        $$f_{\nu}= \frac{(1+z)\,L_{\nu^{\prime}}}{4\,\pi\,d_L^2}.$$
The factor of $(1 + z) $ accounts for
the fact that the flux and luminosity are not bolometric, but are
densities per unit frequency. (The factor would appear in the
denominator if the expression related flux and luminosity densities
per unit wavelength.)

The monochromatic luminosity is therefore
        $$L_{\nu^{\prime}}= \frac{4\,\pi\,d_L^2}{(1+z)}\,f_{\nu}
 =\frac{4\,\pi\,d_L^2}{(1+z)}\,f_{\nu^{\prime}}\Big(\frac{f_{\nu}}{f_{\nu^{\prime}}}\Big)$$ 
but since $f_{\nu}\propto\nu^{\alpha}$ and $\alpha=(1-\Gamma)$,
  $$\Big(\frac{f_{\nu}}{f_{\nu^{\prime}}}\Big)= \Big(\frac{\nu}{\nu^{\prime}}\Big)^{(1-\Gamma)}$$ 
so that
  $$L_{\nu^{\prime}}= 4\,\pi\,d_L^2\,(1+z)^{(\Gamma-2)}\,f_{\nu^{\prime}}$$

\noindent in erg sec$^{-−1}$ Hz$^{-−1}$.   In this way, the flux measured
at $\nu$ in the observed frame yields the monochromatic 
luminosity $L_{\nu^{\prime}}$ in the rest frame. 

The broadband luminosity in erg sec$^{-1}$ is therefore
$$ L_X =  \int_{\nu_1}^{\nu_2}\,L(\nu)\,d\nu = 
4\,\pi\,d_L^2(1+z)^{(\Gamma-2)}\, \int_{\nu_1}^{\nu_2}\,f(\nu)\,d\nu.$$
Then, since
 $$ \int_{\nu_1}^{\nu_2}\,f(\nu)\,d\nu 
= f(\nu_0)\int_{\nu_1}^{\nu_2}\,\Big(\frac{\nu}{\nu_0}\Big)^{(1-\Gamma)}\,d\nu,$$

$$=~\frac{f(\nu_0)}{\nu_0^{(1-\Gamma)}}
\,\Big[\frac{\nu^{(2-\Gamma)}}{(2-\Gamma)}\Big]^{\nu_2}_{\nu_1}
=~\frac{f(\nu_0)}{\nu_0^{(1-\Gamma)}\,}
\frac{\Big[\nu_2^{(2-\Gamma)} - \nu_1^{(2-\Gamma)} \Big]}{(2-\Gamma)}$$
we get
$$ L_X =
4\,\pi\,d_L^2(1+z)^{(\Gamma-2)}\,\frac{f(\nu_0)}{\nu_0^{(1-\Gamma)}}
\,\Big[\frac{\nu_2^{(2-\Gamma)} - \nu_1^{(2-\Gamma)}}{(2-\Gamma)}\Big]$$
where the final term is convenient  for L'Hospital's rule.
Perhaps more intuitively, we can write
$$ L_X =
4\,\pi\,d_L^2\,(1+z)^{(\Gamma-2)}\,
  \Big[\frac{\nu_2\,f(\nu_2) - \nu_1\,f(\nu_1)}{(2-\Gamma)}\Big].$$  
To substitute $E$ in keV for frequencies above,
just multiply by $2.41666\times 10^{17}$ Hz/keV.

%% Put the bibliography here, most people will use BiBTeX in
%% which case the environment below should be replaced with
%% the \bibliography{} command.
% 

\clearpage

\begin{figure*}
\begin{center}
\begin{minipage}[c]{8cm}
        \includegraphics[width=1.0\textwidth]{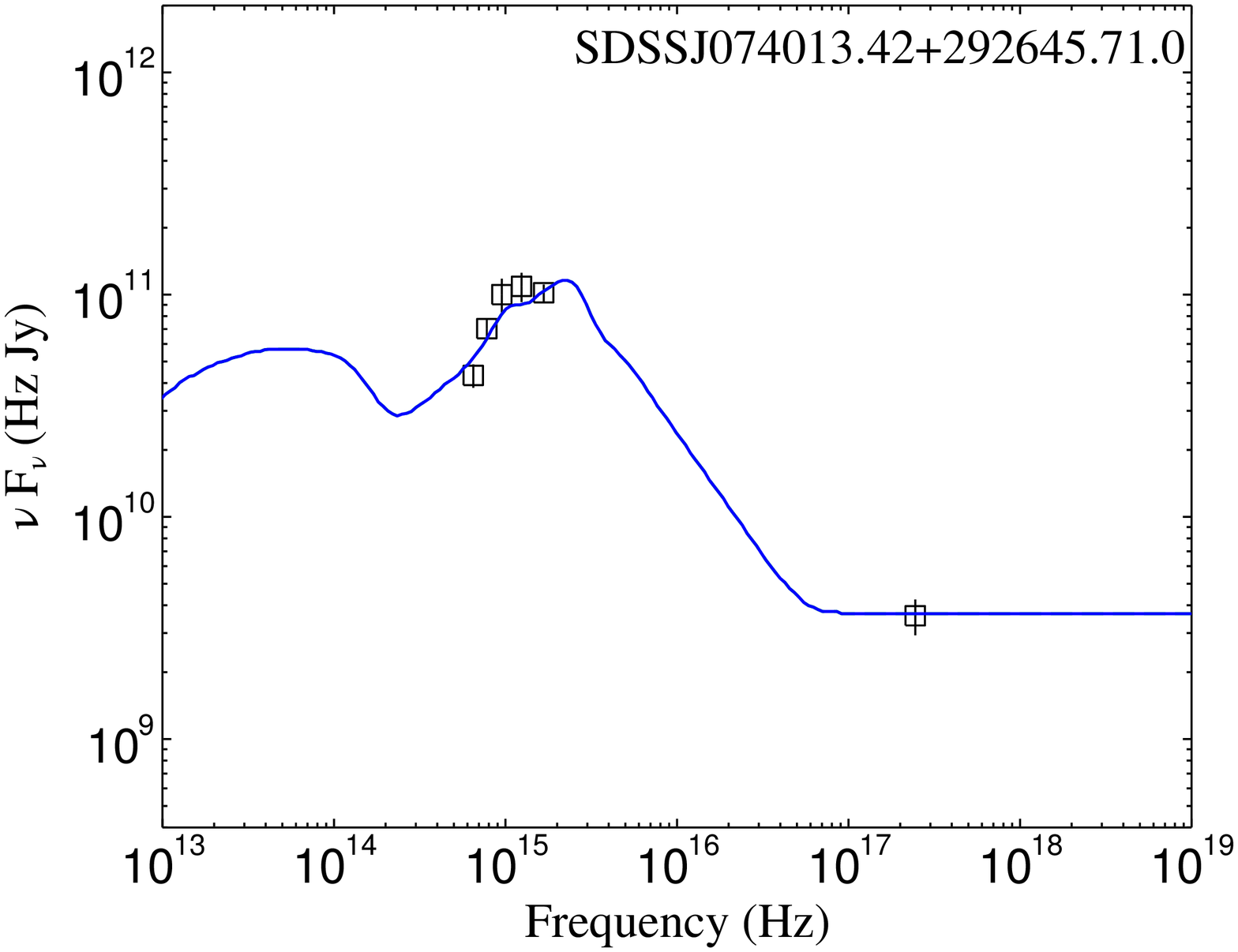}
\end{minipage}
\begin{minipage}[c]{8cm}
        \includegraphics[width=1.0\textwidth]{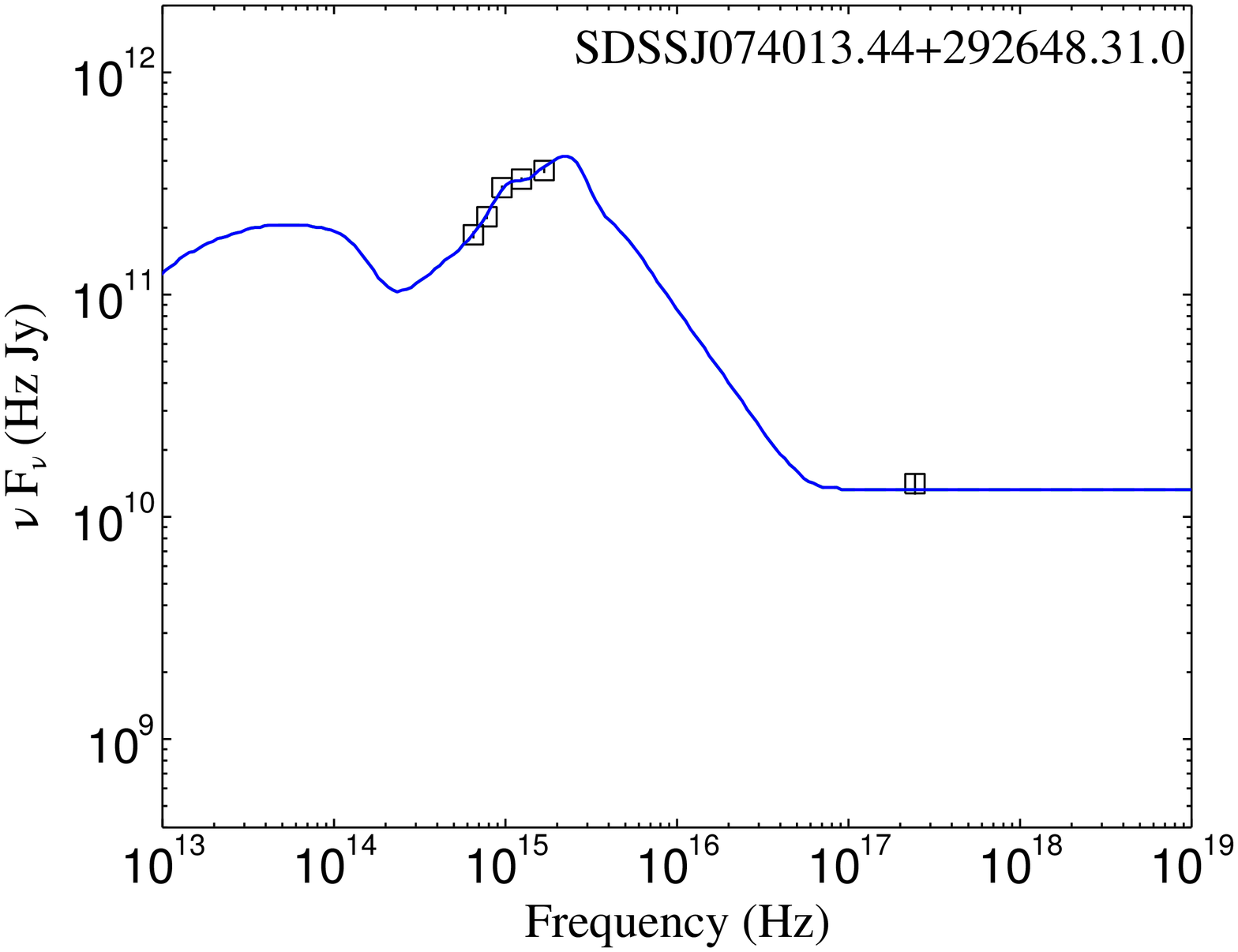}
\end{minipage}
\\
\begin{minipage}[c]{8cm}
        \includegraphics[width=1.0\textwidth]{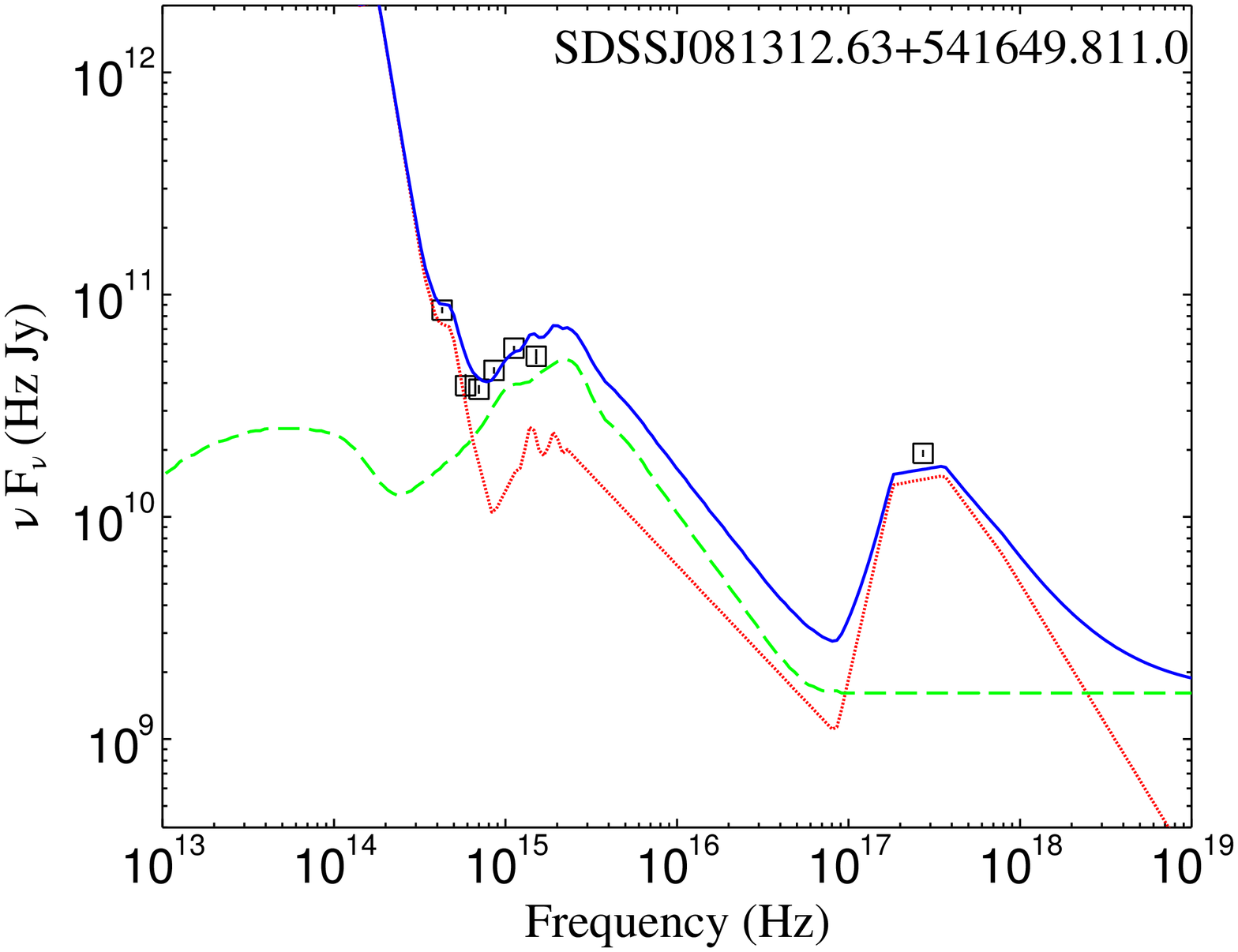}
\end{minipage}
\begin{minipage}[c]{8cm}
        \includegraphics[width=1.0\textwidth]{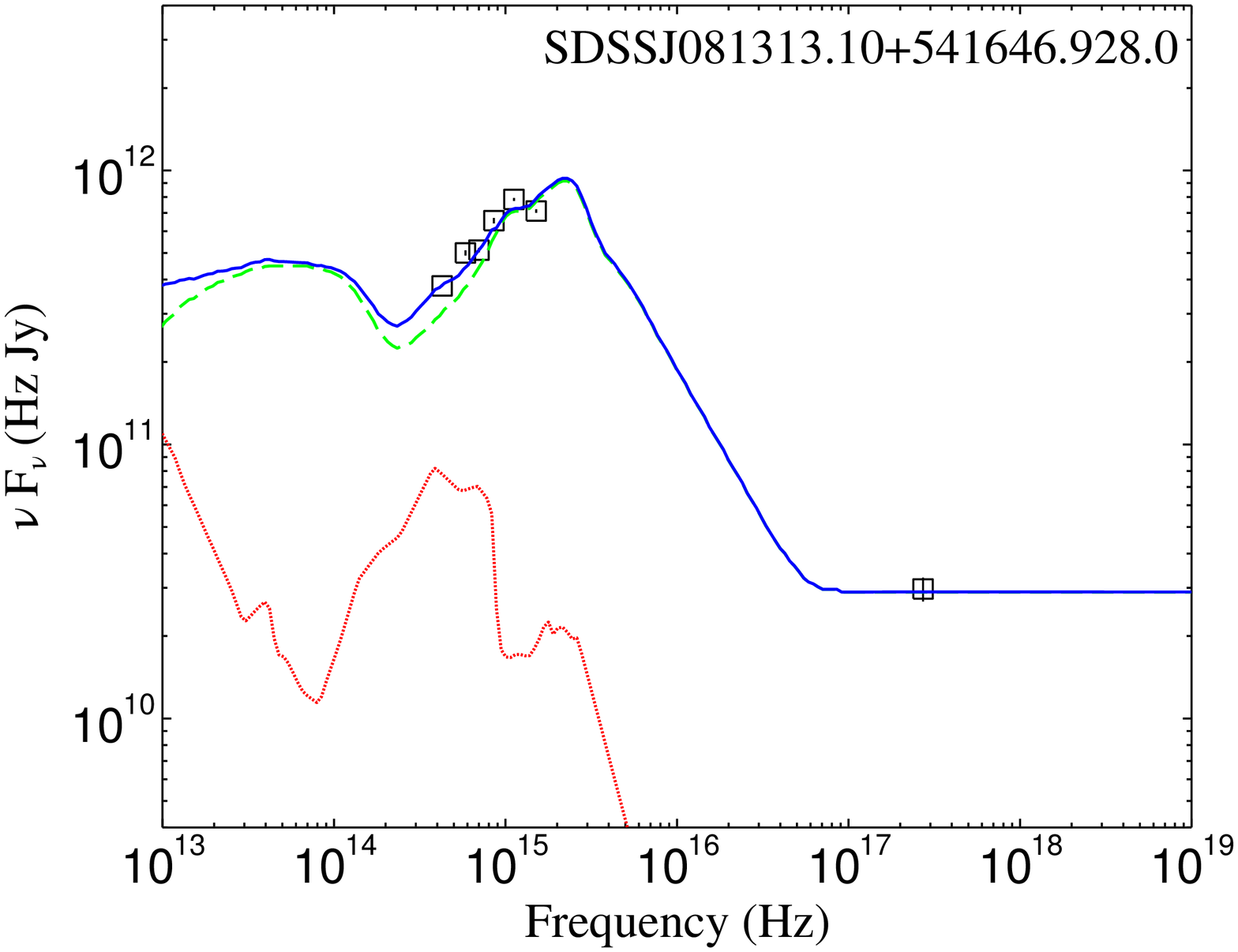}
\end{minipage}
\\
\begin{minipage}[c]{8cm}
        \includegraphics[width=1.0\textwidth]{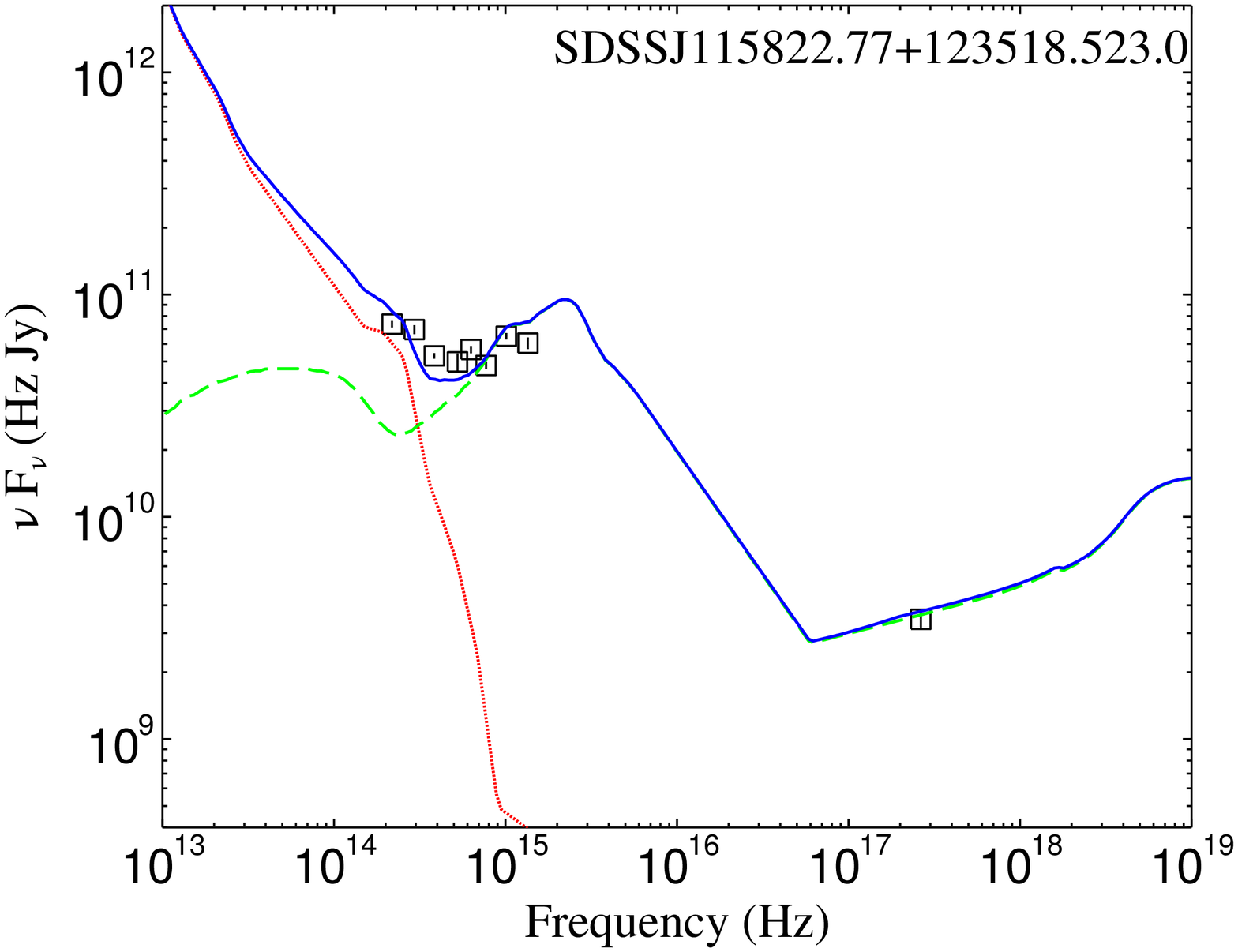}
\end{minipage}
\begin{minipage}[c]{8cm}
        \includegraphics[width=1.0\textwidth]{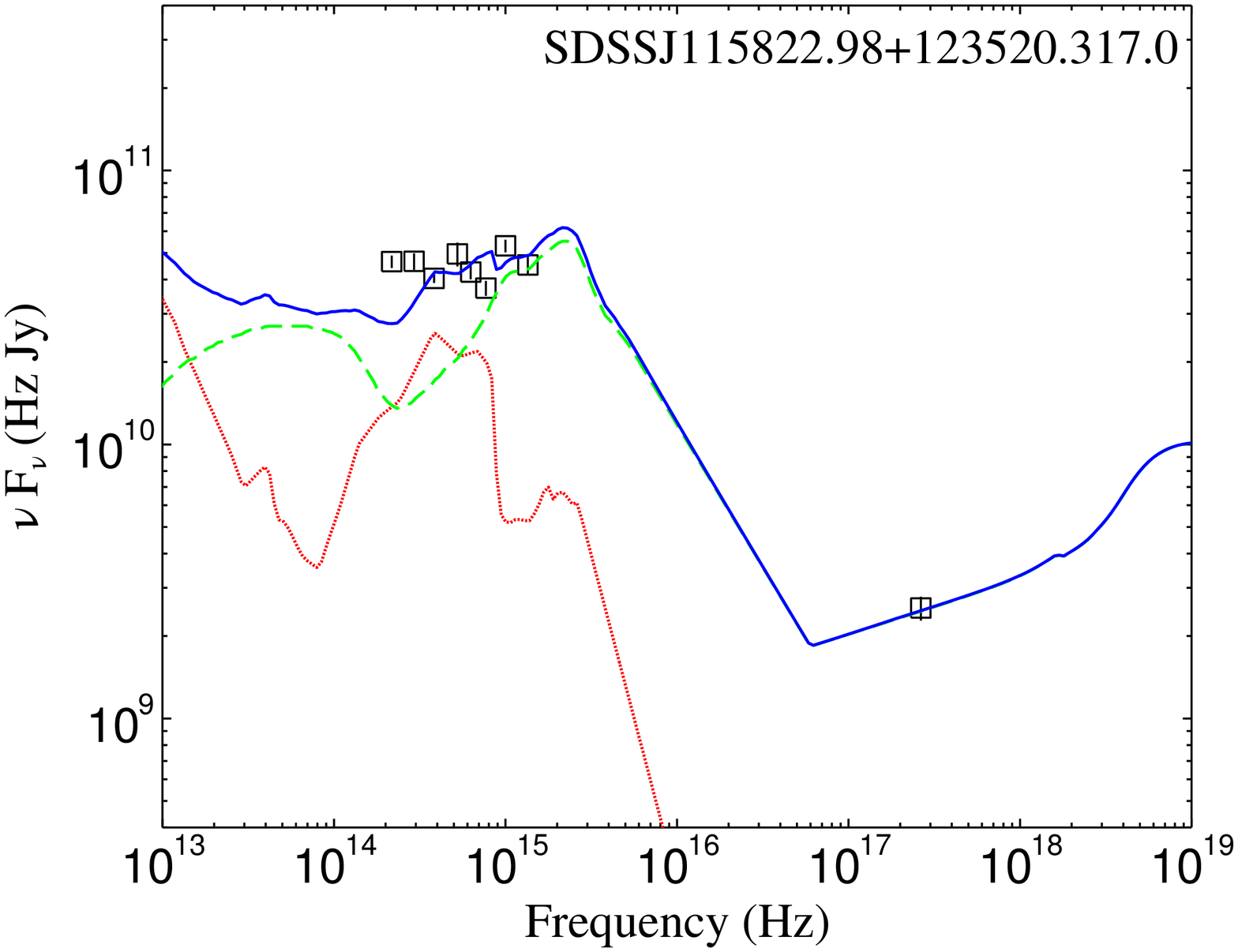}
\end{minipage}
    \caption{Near-infrared to X-ray SEDs (Ruiz et al. 2010) in $\nu
      f_\nu$ for each of our quasar pairs.  Solid blue lines show the total
      predicted SED. Green and red lines are the corresponding AGN
      and starburst templates used, respectively. Parameters for model
      fits are given in Table~\ref{tseds}.} 
\end{center}
\end{figure*}

\clearpage
\begin{figure*}
\begin{center}
\ContinuedFloat
\begin{minipage}[c]{8cm}
        \includegraphics[width=1.0\textwidth]{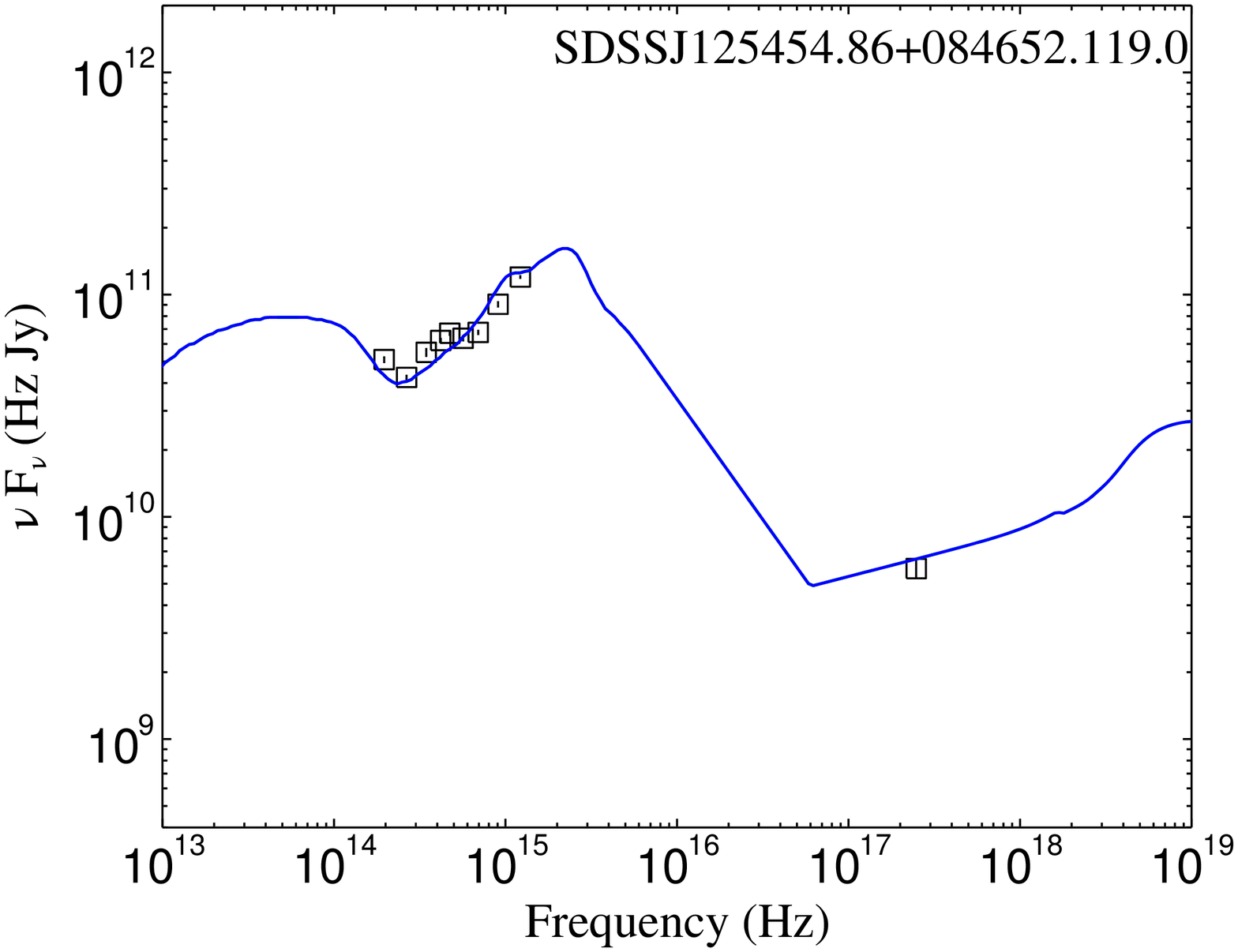}
\end{minipage}
\begin{minipage}[c]{8cm}
        \includegraphics[width=1.0\textwidth]{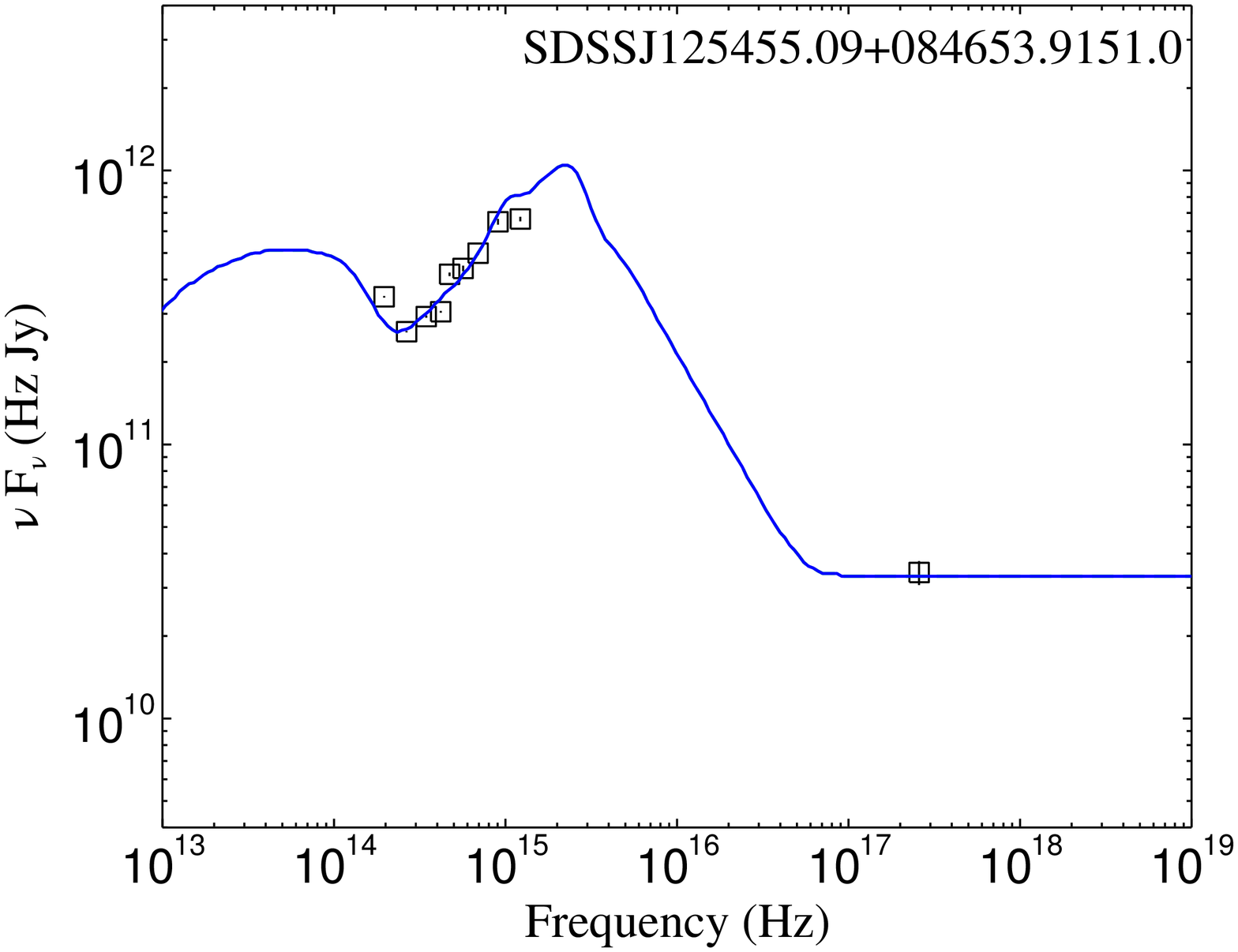}
\end{minipage}
\\
\begin{minipage}[c]{8cm}
        \includegraphics[width=1.0\textwidth]{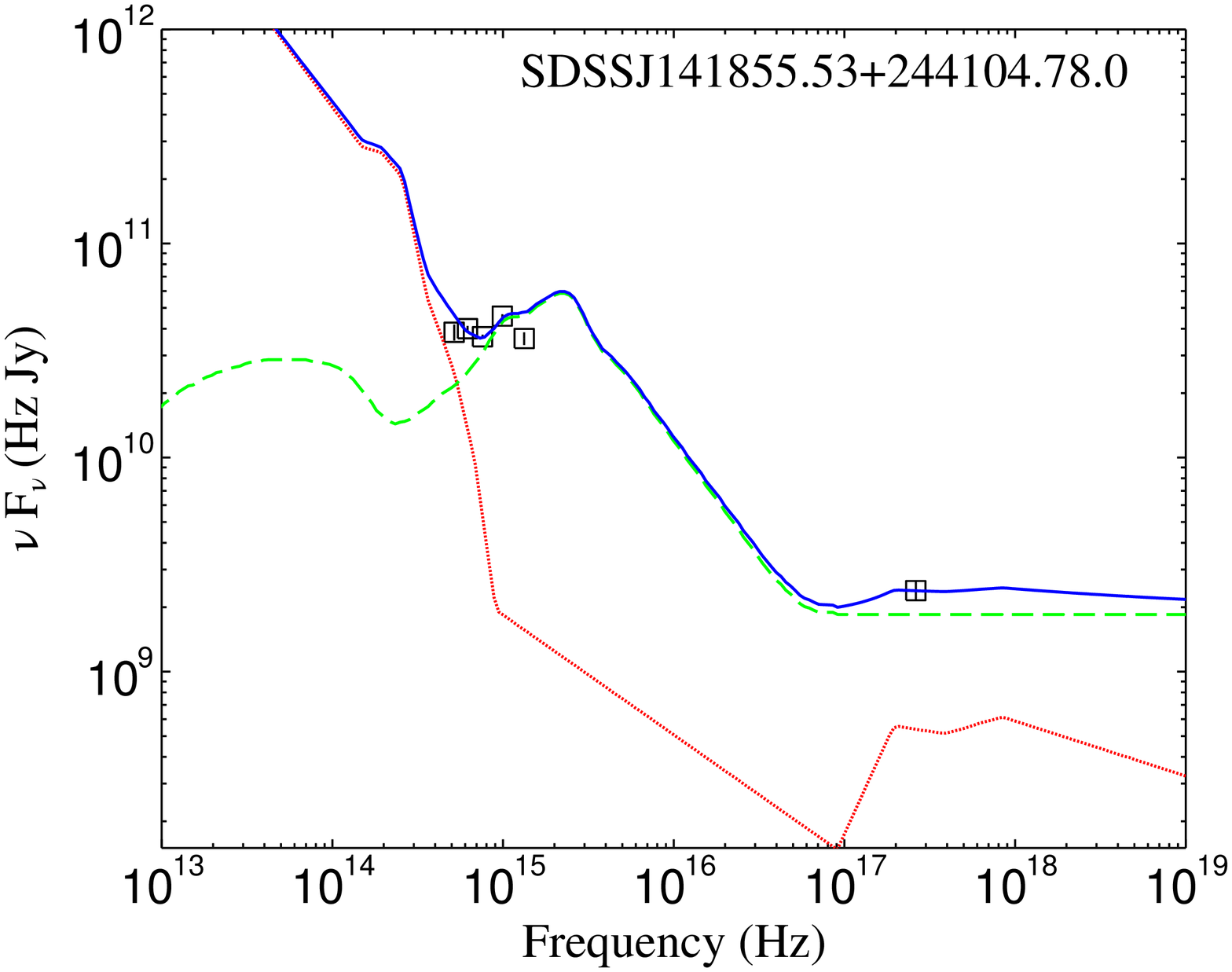}
\end{minipage}
\begin{minipage}[c]{8cm}
        \includegraphics[width=1.0\textwidth]{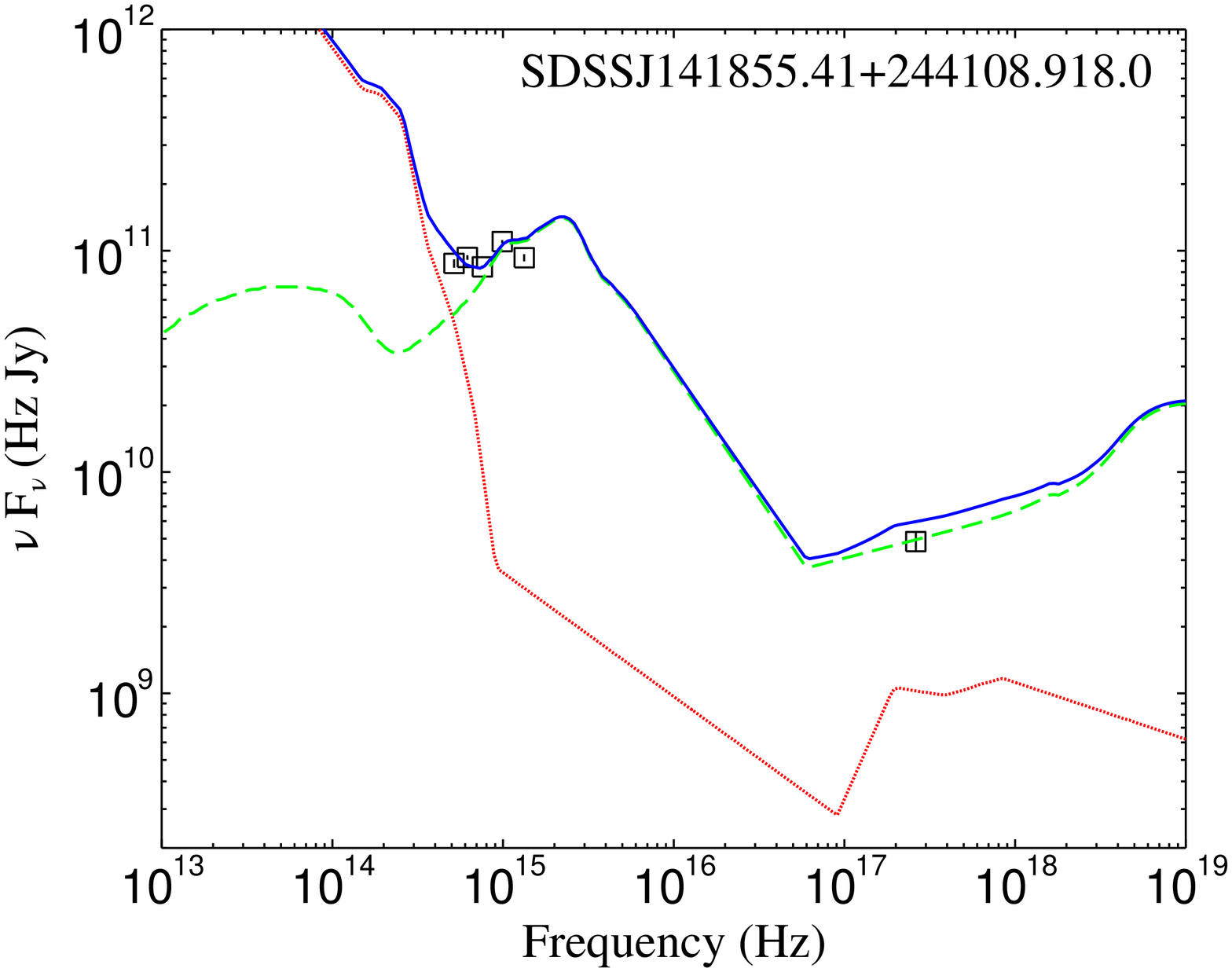}
\end{minipage}
\\
\begin{minipage}[c]{8cm}
        \includegraphics[width=1.0\textwidth]{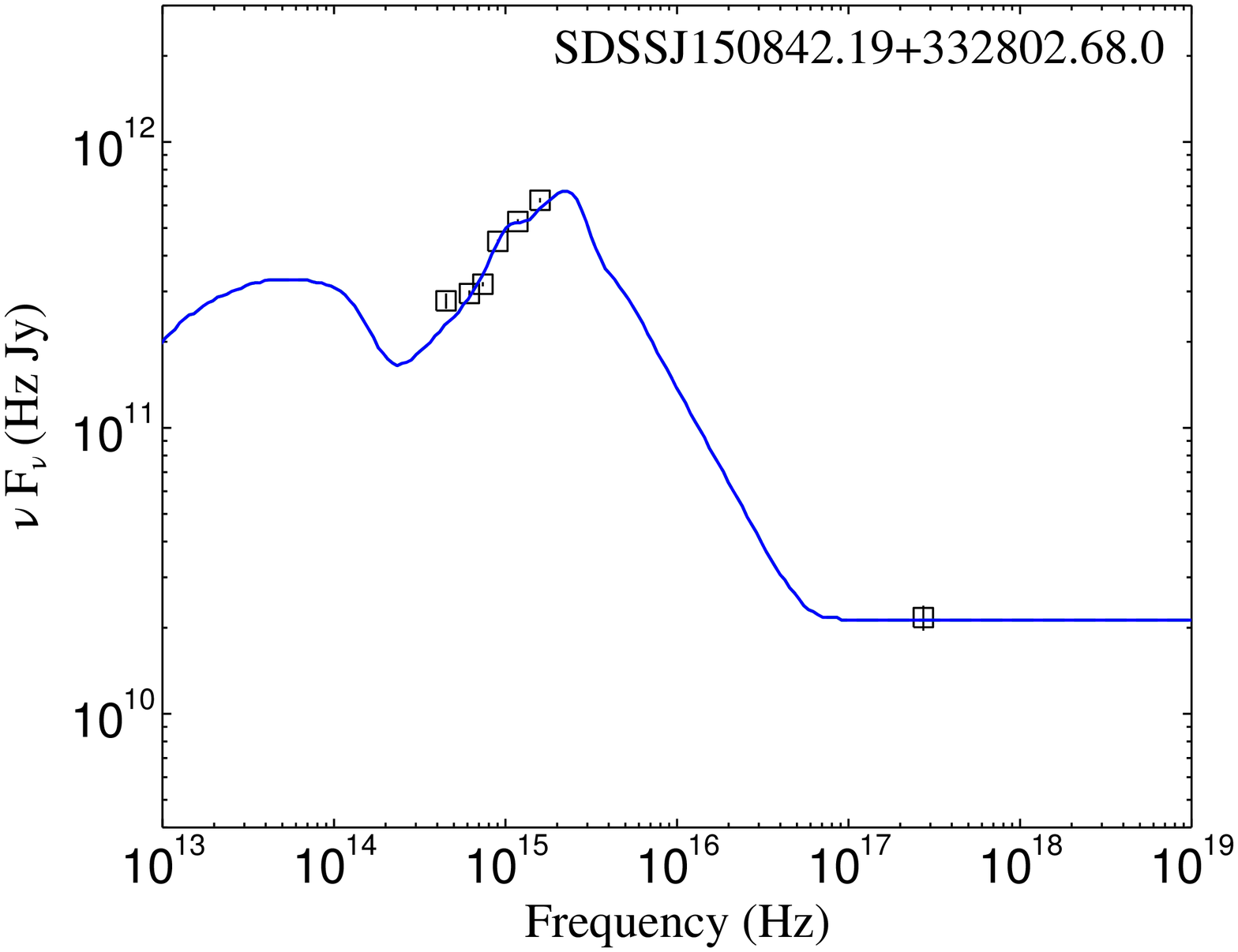}
\end{minipage}
\begin{minipage}[c]{8cm}
        \includegraphics[width=1.0\textwidth]{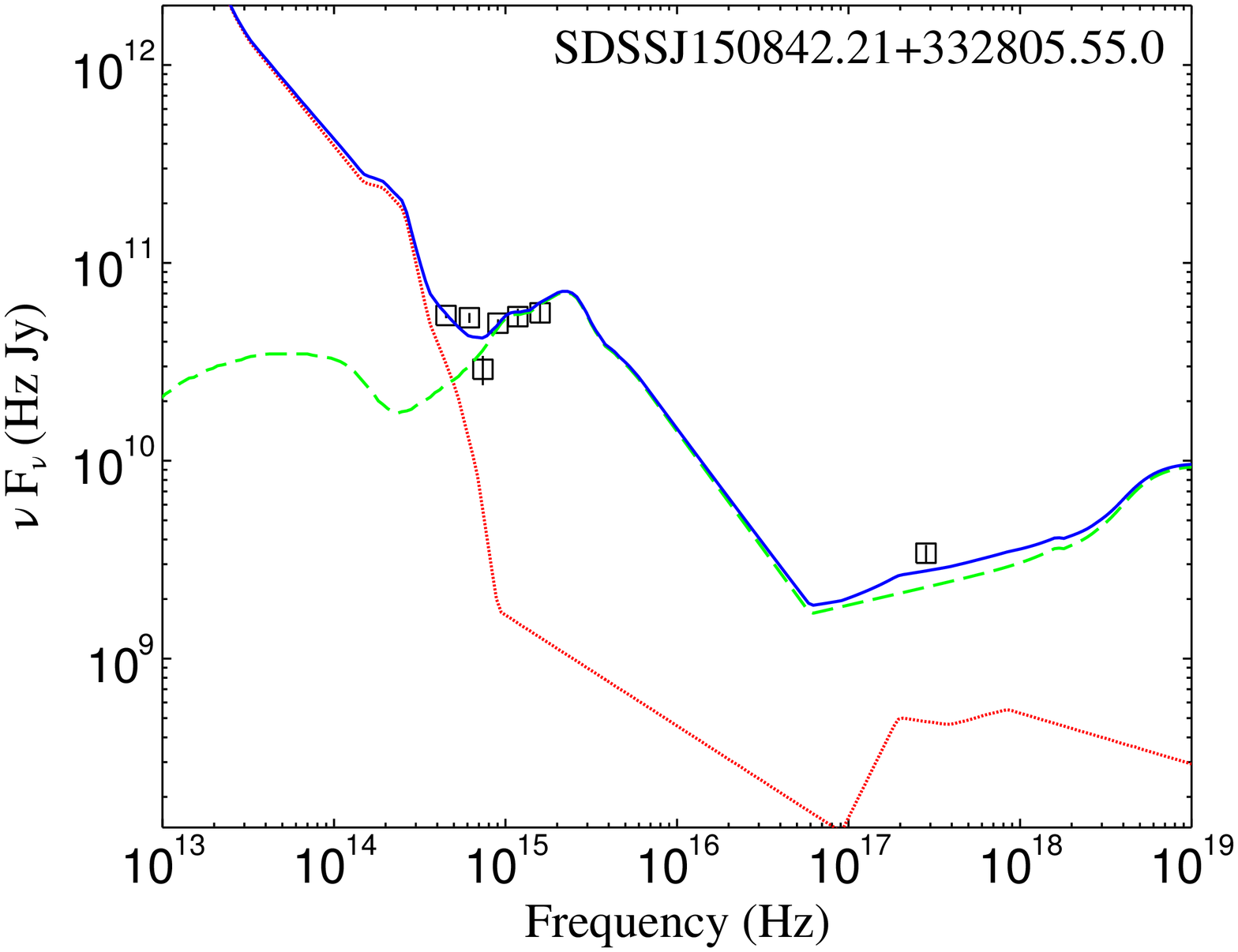}
\end{minipage}
    \caption{Near-infrared to X-ray SEDs (Ruiz et al. 2010) in $\nu
      f_\nu$ for each of our quasar pairs.  Solid blue lines show the total
      predicted SED. Green and red lines are the corresponding AGN
      and starburst templates used, respectively. Parameters for model
      fits are given in Table~\ref{tseds}.} 
\end{center}
\end{figure*}

\clearpage
\begin{figure*}
\begin{center}
\ContinuedFloat
\begin{minipage}[c]{8cm}
        \includegraphics[width=1.0\textwidth]{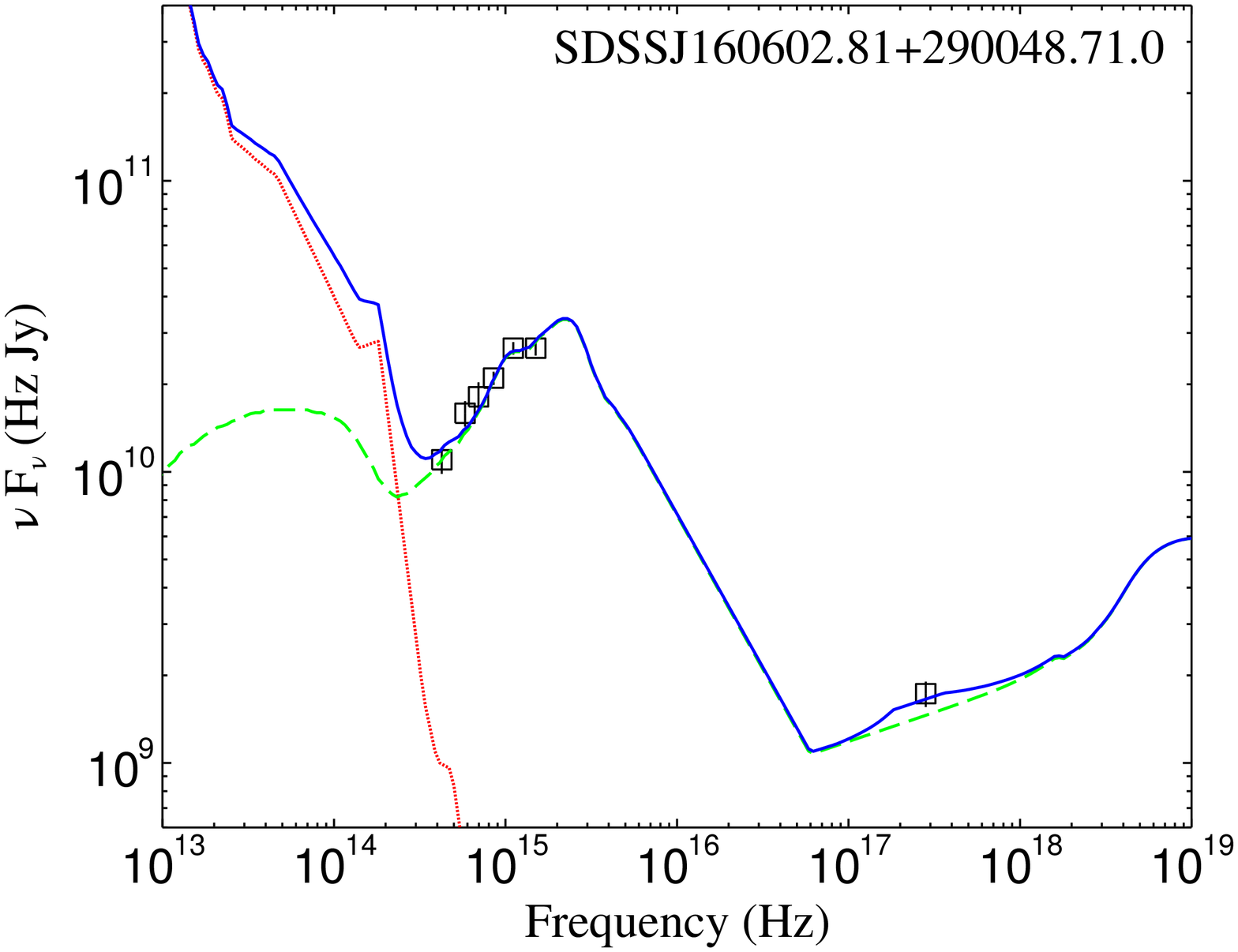}
\end{minipage}
\begin{minipage}[c]{8cm}
        \includegraphics[width=1.0\textwidth]{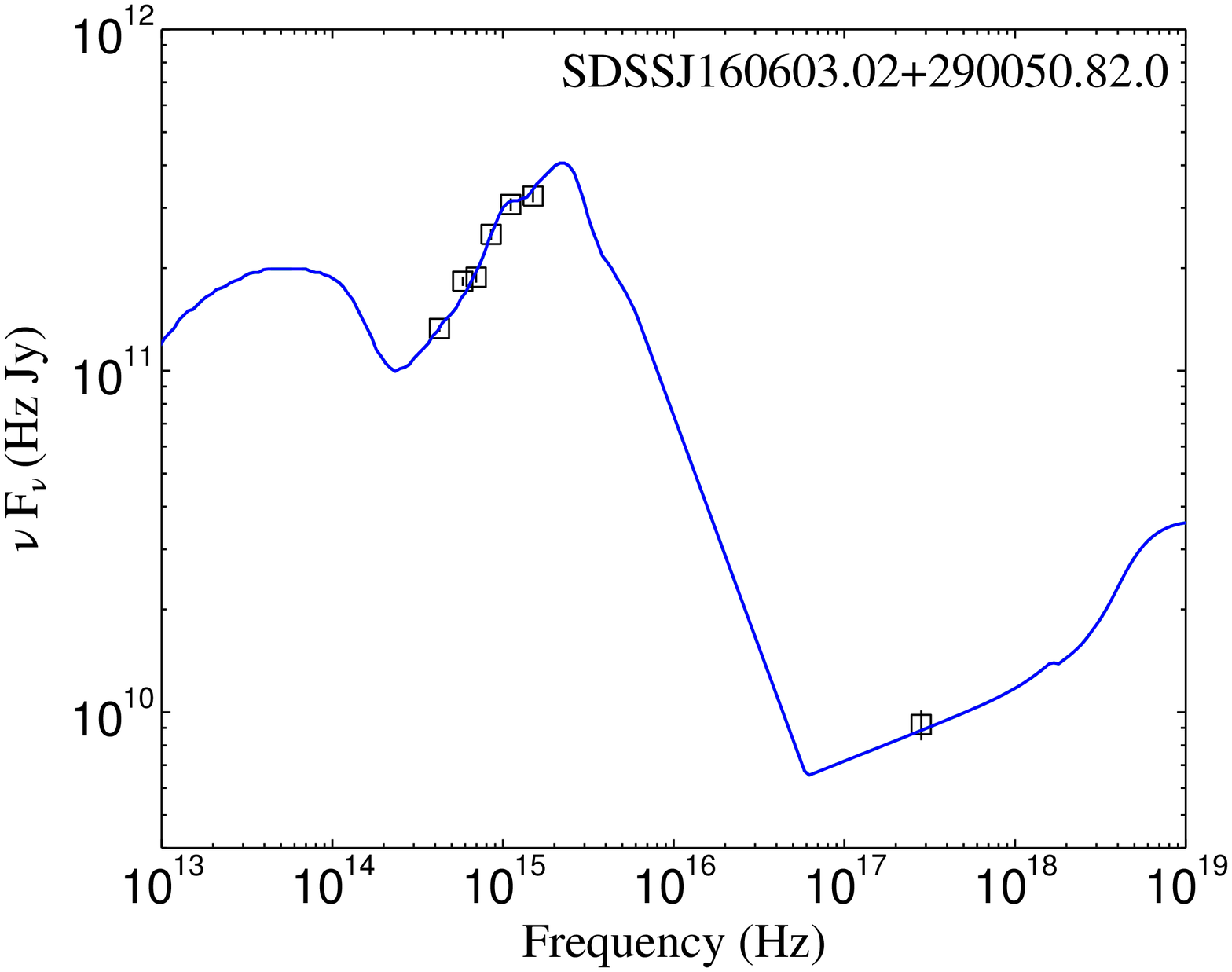}
\end{minipage}
\end{center}
    \caption{Near-infrared to X-ray SEDs (Ruiz et al. 2010) in $\nu
      f_\nu$ for each of our quasar pairs.  Solid blue lines show the total
      predicted SED. Green and red lines are the corresponding AGN
      and starburst templates used, respectively. Parameters for model
      fits are given in Table~\ref{tseds}.} 
    \label{fig:seds1}
\end{figure*} 
\clearpage

% gamma_aox.ps          Gamma looks a bit soft, could be the Nh fits
% gamma_mydeltaaox.ps   Requires too much explaining
% opteml_aox0.ps         

% convert -density 250 SDSSJ0740_histogram2.eps SDSSJ0740_histogram2.jpg
% convert -density 250 SDSSJ0740_variance2.eps SDSSJ0740_variance2.jpg
% montage -geometry +1+1 -density 128 SDSSJ0740_histogram2.jpg SDSSJ0740_variance2.jpg SDSSJ0740fig2.eps

\clearpage
\pagestyle{empty}
\setlength{\voffset}{21mm}

\begin{deluxetable}{cccccc}
\tablecaption{\Chandra\ Close Binary Quasar Sample  \label{tsample}}
\tabletypesize{\small}
\tablewidth{0pt}
\tablehead{
\colhead{Pair Name} & \colhead{ObsID} &\colhead{Exposure}
&\colhead{$\theta$ \tablenotemark{a}} &\colhead{$R_p$ \tablenotemark{b}}
&\colhead{Galactic $\mbox{N}_{\mbox{H}}$\tablenotemark{c} }\\
 &  & \colhead{(sec)}  & \colhead{(arcsec)} & \colhead{(kpc)}
& \colhead{($10^{20}~\mbox{cm}^{-2}$)} }
\startdata
% QSO Pair	ObsId	Exposure theta  Rp    NhGal
SDSS\,J0740+2926 & 10312 & 20859 &  2.6 & 15.0  & 4.24 \\
SDSS\,J0813+5416 & 10313 & 30625 &  5.0 & 26.9  & 4.21 \\
SDSS\,J1158+1235 & 10314 & 30827 &  3.6 & 17.0  & 2.07 \\
SDSS\,J1254+0846 & 10315 & 15967 &  3.8 & 15.4  & 1.92 \\
SDSS\,J1418+2441 & 10316 & 29762 &  4.5 & 21.0  & 2.00 \\
SDSS\,J1508+3328 & 10317 & 31317 &  2.9 & 16.0  & 1.51 \\
SDSS\,J1606+2900 & 10318 & 12852 &  3.5 & 18.4  & 3.19 \\
\enddata
\tablenotetext{a}{Separation between QSO components in arcsec.}
\tablenotetext{b}{Proper separation between QSO components in kpc.}
\tablenotetext{c}{ Galactic column in units 10$^{20}$\,cm$^{-2}$
from the NRAO dataset of \citet{Dickey90}.}
\end{deluxetable}

\clearpage
\pagestyle{empty}
\setlength{\voffset}{21mm}
% see /data/green/prop/AXAF/wsqps/2008/Spectra/aaanotes
% List within pairs by RA
\begin{landscape}
\begin{deluxetable}{lclrrccrrr}
%\rotate use landscape instead for emulateapj
\tabletypesize{\small}
\tablewidth{0pt}
\tablecaption{Chandra Close Binary Quasar Sample \label{tfits}}
% SDSS   rmag   z  counts GPL\pm GPLabs\pm fx Lx aox
\tablehead{
 \colhead{SDSS NAME} &\colhead{$r$ mag}\tablenotemark{a}&\colhead{$z$\tablenotemark{b}}
 & \colhead{Counts\tablenotemark{c}} 
 & \colhead{$\Gamma_{PL}$\tablenotemark{d}} 
 &\colhead{$\Gamma_{PLAbs}$\tablenotemark{d}}
 &\colhead{$N_H^z~$\tablenotemark{e}}  
 &\colhead{log\,$f_X$\tablenotemark{f}} &\colhead{log\,$ l_X$\tablenotemark{g}} &\colhead{$\aox$\tablenotemark{h}} } 
%
% LISTED IN RA ORDER.
\startdata
% obsid srcid from Tom
% 10312 2
J074013.42+292645.7 & 19.47 &  0.978~ H &  82.4 &
   1.85$^{+0.20}_{-0.20}$& 1.85$^{+0.26}_{-0.20}$ & $<$27 
   & -13.527 & 25.993 & 1.583 \\
% 10312 1
J074013.44+292648.3 & 18.27 & 0.9803 S &   288.5 &
   2.24$^{+0.11}_{-0.11}$ &
   2.24$^{+0.11}_{-0.11}$ & $<$9 
  & -13.044 & 26.589 & 1.540 \\
\multicolumn{10}{c}{}\\
% 10313 1
J081312.63+541649.8 & 20.08 & 0.7814 K4m &  195.7 &
    2.51$^{+0.15}_{-0.14}$ & 
    2.51$^{+0.15}_{-0.14}$ & $<$12
    & -13.414 & 26.006 & 1.426 \\
% 10313 2 
J081313.10+541646.9 & 17.18 & 0.7792 S &  2336.8 &
    1.90$^{+0.06}_{-0.06}$ &
    1.93$^{+0.04}_{-0.04}$ & $<$1
    & -12.255 & 27.051 & 1.460 \\

\multicolumn{10}{c}{}\\                                                                      
% 10314 2 
J115822.77+123518.5 & 19.85 & 0.5996 M08  & 367.7 & 
    2.26$^{+0.10}_{-0.10}$ &
    2.51$^{+0.19}_{-0.18}$ & 18$^{+11}_{-11}$ 
    & -13.098 & 26.017 & 1.335 \\
% 10314 1 
J115822.98+123520.3 & 20.12 & 0.5957 M08 & 413.6 & 
    2.16$^{+0.09}_{-0.09}$ &
    2.14$^{+0.16}_{-0.14}$ & $<$9
    & -13.069 & 25.999 & 1.292 \\
\multicolumn{10}{c}{}\\
% 10315 2
J125454.86+084652.1 & 19.43 & 0.4401 G &   349.5 & 
    2.11$^{+0.10}_{-0.10}$ &
    2.11$^{+0.10}_{-0.10}$ &  $<$2
    & -12.853 & 25.887 & 1.355 \\
% 10315 1
J125455.09+084653.9 & 17.08 & 0.4392 G & 1795.5 &
    2.04$^{+0.04}_{-0.04}$ &
    2.04$^{+0.04}_{-0.04}$ & $<$1
    & -12.129 & 26.597 & 1.431 \\
\multicolumn{10}{c}{}\\
% 10316 1
J141855.41+244108.9 & 19.21 & 0.5728 S & 864.9 & 
   1.94$^{+0.06}_{-0.06}$ & 
   1.99$^{+0.11}_{-0.10}$ & $<$7
   & -12.692 & 26.305 & 1.282 \\
% 10316 2
J141855.53+244104.7 & 20.13 & 0.5751 M08 &  33.0 & 
   0.72$^{+0.28}_{-0.28}$ &
   1.6$^{+0.60}_{-0.54}$  &  158$^{+113}_{-87}$ 
   & -13.769 & 25.132 & 1.575 \\
\multicolumn{10}{c}{}\\
% 10317 2
J150842.19+332802.6 & 17.80 & 0.8773 S & 1040.9 &
   2.10$^{+0.06}_{-0.06}$ &
   2.10$^{+0.06}_{-0.06}$ & $<$8
   & -12.670 & 26.807 & 1.514 \\
% 10317 1
J150842.21+332805.5 & 20.19 & 0.878~ H & 111.1  & 
   2.37$^{+0.45}_{-0.40}$ &
   2.61$^{+0.37}_{-0.20}$ & $<$27
   & -13.708 & 25.860 & 1.479 \\
\multicolumn{10}{c}{}\\
% 10318 2
J160602.81+290048.7 & 18.35 & 0.7701 S & 6.1 & 
   0.60$^{+0.71}_{-0.73}$ & 
   2.1 &  $<$0.01
   & -14.491 & 24.844 & 2.129 \\
% 10318 1 
J160603.02+290050.8 & 18.25 & 0.7692 M08 &  144.1 & 
   2.20$^{+0.43}_{-0.42}$ &
   2.44$^{+0.17}_{-0.16}$ & $<$6  
   & -13.171 & 26.222 & 1.611 \\
\enddata
\tablenotetext{a}{SDSS dereddened PSF magnitude.}
\tablenotetext{b}{Redshift. K4m -  February, 2008 KPNO/4m; M08 - Myers
  et al. (2008); H - Hennawi et al. (2006); G - Green et al. (2010); S
  - SDSS } 
\tablenotetext{c}{Net 0.5-8~keV counts.}
\tablenotetext{d}{Best-fit X-ray power-law photon index.
  Uncertainties are the 68\% confidence limits.  If no uncertainties
  are shown, then the value is frozen to enable fitting of $N_H^z$.
  For J160602.81+290048.7, we freeze $\Gamma$ simply to fit the
  overall normalization.} 
\tablenotetext{e}{Best-fit intrinsic column for $PLAbs$ model in units
  10$^{20}$\,cm$^{-2}$.  Upper limits are at 68\% confidence.}
\tablenotetext{f}{X-ray flux (0.5-8\,keV) in erg~cm$^{-2}$~s$^{-1}$
calculated using the $PLAbs$ model.}
\tablenotetext{g}{X-ray luminosity at 2\,keV in  erg~s$^{-1}$~Hz$^{-1}$.}
\tablenotetext{h}{\aox, the optical/UV to X-ray spectral index.}
\end{deluxetable}
\clearpage
\end{landscape}
   % X-ray fit results
\setlength{\voffset}{0mm}

\pagestyle{empty}
\begin{landscape}
\begin{deluxetable}{cccccccccccc}
%\rotate use landscape instead for emulateapj
\tabletypesize{\small}
\tablewidth{0pt}
\tablecaption{Chandra Binary Quasar Sample Near-Infrared Properties 
\label{tnir}}
\tablehead{
\colhead{SDSS Name}
&\colhead{Exposure \tablenotemark{a}} 	
&\colhead{$J_{S}$ \tablenotemark{b}} 	
&\colhead{err$J_{S}$ \tablenotemark{c}}
&\colhead{$J_{U}$ \tablenotemark{d}} 
&\colhead{err$J_{U}$ \tablenotemark{e}}
&\colhead{Y \tablenotemark{f}} 
&\colhead{errY \tablenotemark{g}}
&\colhead{H \tablenotemark{h}} 
&\colhead{errH \tablenotemark{i}}
&\colhead{K \tablenotemark{j}} 
&\colhead{errK \tablenotemark{k}}}
\startdata
J074013.42+292645.7 &\ldots&\ldots&\ldots&\ldots&\ldots&\ldots&\ldots&\ldots&\ldots&\ldots&\ldots \\
J074013.44+292648.3 &\ldots &\ldots&\ldots&\ldots&\ldots&\ldots&\ldots&\ldots&\ldots&\ldots&\ldots\\\\
J081312.63+541649.8 &2610&17.869&0.034&\ldots&\ldots&\ldots&\ldots&\ldots&\ldots&\ldots&\ldots\\
J081313.10+541646.9 &2610&18.744&0.061&\ldots&\ldots&\ldots&\ldots&\ldots&\ldots&\ldots&\ldots\\\\
J115822.77+123518.5 &900&17.174&0.013&\ldots&\ldots&\ldots&\ldots&17.085&0.050&16.239&0.036\\
J115822.98+123520.3 &900&17.411&0.016&\ldots&\ldots&\ldots&\ldots&17.515&0.074&16.732&0.056\\\\
J125454.86+084652.1 &1260 &18.098&0.061&17.925&0.045&18.424&0.034&17.066&0.022&16.129&0.029\\
J125455.09+084653.9 &1260 &16.063&0.010&16.079&0.009&16.505&0.008&15.433&0.006&14.327&0.007\\\\
J141855.41+244108.9 &\ldots&\ldots&\ldots&\ldots&\ldots&\ldots&\ldots&\ldots&\ldots&\ldots&\ldots\\
J141855.53+244104.7 &\ldots&\ldots&\ldots&\ldots&\ldots&\ldots&\ldots&\ldots&\ldots&\ldots&\ldots\\\\
J150842.19+332802.6 & 630 &16.551&0.058&16.629&0.012&\ldots&\ldots&\ldots&\ldots&\ldots&\ldots\\
J150842.21+332805.5 &630&18.473&0.031&18.463&0.063&\ldots&\ldots&\ldots&\ldots&\ldots&\ldots\\\\
J160602.81+290048.7 &\ldots&\ldots&\ldots&17.398&0.023&\ldots&\ldots&\ldots&\ldots&\ldots&\ldots\\
J160603.02+290050.8 &\ldots&\ldots&\ldots&17.369&0.022&\ldots&\ldots&\ldots&\ldots&\ldots&\ldots\\
\enddata
\tablenotetext{a}{Total MMT-SWIRC exposure time in seconds.}
\tablenotetext{b}{SWIRC J-band magnitudes.} 	
\tablenotetext{c}{Error in SWIRC J-band magnitudes.}
\tablenotetext{d}{UKIDSS J-band magnitudes.} 
\tablenotetext{e}{Error in UKIDSS J-band magnitudes.}
\tablenotetext{f}{UKIDSS Y-band magnitudes.} 
\tablenotetext{g}{Error in UKIDSS Y-band magnitudes.}
\tablenotetext{h}{UKIDSS H-band magnitudes.} 
\tablenotetext{i}{Error in UKIDSS H-band magnitudes.}
\tablenotetext{j}{UKIDSS K-band magnitudes.} 
\tablenotetext{k}{Error in UKIDSS K-band magnitudes.}
\end{deluxetable}
\clearpage
\end{landscape}
  % MMT/SWIRC NIR table

\begin{table*}
\caption{Quasar Spectral Energy Distribution Fit Results}
\centering
\begin{minipage}{10in}
\begin{tabular} {cccccc}
\hline\hline
SDSS Name & \multicolumn{1}{c}{log\,L$_{Bol}$\tablenotemark{a}} 
& $J_{AGN}$\tablenotemark{b}  
& $J_{SB}$\tablenotemark{c} 
& $P_{IRX}$\tablenotemark{d} 
& $P_{Bol}$\tablenotemark{e}\\ 
\hline\hline
J074013.42+292645.7 & 46.362 & QSO   & \ldots          &100/0  & 100/0 \\
J074013.44+292648.3 & 46.909 & QSO   & \ldots          &100/0  & 100/0\\\\
J081312.63+541649.8 & 48.159 & QSO   & M82             &71/29 & $<$10/$>$90\\
J081313.10+541646.9 & 47.087 & QSO   & NGC7714         &79/21 & 79/21\\\\
J115822.77+123518.5 & 47.185 & LDQSO & IRAS 12112+0305 &74/26 & $<$10/$>$90\\
J115822.98+123520.3 & 45.770 & LDQSO & NGC7714         &58/42 & 58/42\\\\
J125454.86+084652.1 & 45.787 & LDQSO & \ldots          &100/0  & 100/0\\
J125455.09+084653.9 & 46.457 & QSO   & \ldots          &100/0  & 100/0\\\\
J141855.41+244108.9 & 48.009 & LDQSO & IRAS 12112+0305 &80/20 & $<$10/$>$90\\  
J141855.53+244104.7 & 47.713 & QSO   & IRAS 12112+0305 &76/24 & $<$10/$>$90\\\\
J150842.19+332802.6 & 47.013 & QSO   & \ldots          &100/0  & 100/0\\
J150842.21+332805.5 & 48.127 & LDQSO & IRAS 12112+0305 &77/23 & $<$10/$>$90\\\\
J160602.81+290048.7 & 48.225 & LDQSO & M82             &68/32 & $<$10/$>$90\\
J160603.02+290050.8 & 46.641 & LDQSO & \ldots          &100/0  & 100/0\\
\hline\hline
\end{tabular}
\end{minipage}
\label{tseds}
\tablenotetext{a}{Log of the luminosity from 10$^{9.}$ - 10$^{19.}$ Hz
  in units of erg~s$^{-1.}$ from  template fit.}
\tablenotetext{b}{AGN template used in the best fit solution: QSO
  (radio quiet QSO), LDQSO (luminosity-dependent QSO template).}
\tablenotetext{c}{Starburst template used in the best fit.}
\tablenotetext{d}{Percent QSO/starburst contribution in the range
  10$^{14.}$ - 10$^{18.}$ Hz.}
\tablenotetext{e}{Percent QSO/starburst contribution in the range 
  10$^{9.}$ - 10$^{19.}$ Hz.}
\end{table*}

\end{document}